\documentclass[acmsmall]{acmart}
\AtBeginDocument{%
  }

\usepackage[resetlabels]{multibib}
\newcites{game}{Ludography}
\newcommand{\citegameprefix}{G}
\usepackage{xparse}
\let\origcitegame\citegame
\RenewDocumentCommand{\citegame}{O{} O{} m}{%
  \renewcommand{\citenumfont}[1]{\citegameprefix##1}%  
  \origcitegame[#1][#2]{#3}%
  \renewcommand{\citenumfont}[1]{##1}%
}
\setcopyright{cc}
\setcctype{by}
\acmDOI{10.1145/3748633}
\acmYear{2025}
\acmJournal{PACMHCI}
\acmVolume{9}
\acmNumber{6}
\acmArticle{GAMES038}
\acmMonth{10}
\received{2025-02-19}
\received[accepted]{2025-07-24}
\begin{document}

\title[Elaborating on the Effects of Eudaimonic Gaming Experiences]{“I Would Not Be This Version of Myself Today”: Elaborating on the Effects of Eudaimonic Gaming Experiences}

\author{Nisha Devasia}
\orcid{0000-0002-1506-5443}
\affiliation{%
  \institution{University of Washington}
  \city{Seattle}
  \country{USA}
}
\email{ndevasia@uw.edu}

\author{Georgia Kenderova}
\orcid{0000-0001-5061-2273}
\affiliation{%
  \institution{University of Washington}
  \city{Seattle}
  \country{USA}
}
\email{gak98@uw.edu}

\author{Michele Newman}
\orcid{0000-0002-5293-7992}
\affiliation{%
  \institution{University of Washington}
  \city{Seattle}
  \country{USA}
}
\email{mmn13@uw.edu}

\author{Julie A. Kientz}
\orcid{0000-0001-7437-7861}
\affiliation{%
  \institution{University of Washington}
  \city{Seattle}
  \country{USA}
}
\email{jkientz@uw.edu}

\author{Jin Ha Lee}
\orcid{0000-0002-9007-514X}
\affiliation{%
  \institution{University of Washington}
  \city{Seattle}
  \country{USA}
}
\email{jinhalee@uw.edu}

\renewcommand{\shortauthors}{Devasia et al.}

\begin{abstract}
While much of the research in digital games has emphasized hedonic experiences, such as flow, enjoyment, and positive affect, recent years have seen increased interest in \textit{eudaimonic} gaming experiences, typically mixed-affect and associated with personal meaningfulness and growth. The formation of such experiences in games is theorized to have four constituent elements: motivation, game use, experience, and effects. However, while the first three elements have been relatively well explored in the literature, the effects - and how they may influence positive individual outcomes - have been underexplored thus far. To this end, in this work, we investigate the perceived outcomes of eudaimonic gaming and how different components of the experience influence these effects. We conducted a survey (n = 166) in which respondents recounted meaningful gaming experiences and how they affected their present lives. We used a mixed-methods approach to classify effects and identify significant subcomponents of their formation. We contribute an empirical understanding of how meaningful gaming experiences can lead to positive reflective, learning, social, health, and career effects, extending current theoretical models of eudaimonic gaming experiences and offering implications for how researchers and practitioners might use these findings to promote positive outcomes for players.  
\end{abstract}

\begin{CCSXML}
<ccs2012>
<concept>
<concept_id>10003120.10003121.10003126</concept_id>
<concept_desc>Human-centered computing~HCI theory, concepts and models</concept_desc>
<concept_significance>300</concept_significance>
</concept>
<concept>
<concept_id>10003120.10003121.10011748</concept_id>
<concept_desc>Human-centered computing~Empirical studies in HCI</concept_desc>
<concept_significance>300</concept_significance>
</concept>
<concept>
<concept_id>10010405.10010476.10011187.10011190</concept_id>
<concept_desc>Applied computing~Computer games</concept_desc>
<concept_significance>500</concept_significance>
</concept>
</ccs2012>
\end{CCSXML}

\ccsdesc[300]{Human-centered computing~HCI theory, concepts and models}
\ccsdesc[300]{Human-centered computing~Empirical studies in HCI}
\ccsdesc[500]{Applied computing~Computer games}

\keywords{video games, eudaimonic gaming experiences, player experience, effects of video games}

\maketitle

\section{Introduction}
The idea of \textit{eudaimonia} originated in Aristotelian times \cite{McDowell_2023} and is broadly defined as ``living life in a full and deeply satisfying way'' \cite{Deci_Ryan_2008}. In most two-factor models of media entertainment \cite{vorderer2004enjoyment}, eudaimonia is contrasted with \textit{hedonia}, the former referring to experiences of `appreciation' going beyond mere pleasure \cite{Oliver_Bartsch_2010}, while the latter traditionally refers to general `enjoyment,' i.e., the attempt to elicit a positive affective state \cite{vorderer2011s}. Eudaimonia can be associated with negative affective states; for example, stress, anxiety, and concern about personal expression are associated with lower happiness but higher meaningfulness \cite{Baumeister_Vohs_Aaker_Garbinsky_2015}. This framing, when applied to the study of digital games, primarily concerns player experience. Pleasurable experiences mediated by factors such as flow, positive affect, and other elements of hedonia have been the dominant paradigm in HCI game studies \cite{Cole_Gillies_2022, Mekler_Bopp_Tuch_Opwis_2014, Raney_Oliver_Bartsch_2019, Hamari_Keronen_2017}. However, the field has started to additionally investigate meaningful or eudaimonic experiences in games \cite{Rogers_Woolley_Sherrick_Bowman_Oliver_2017, Daneels_Bowman_Possler_Mekler_2021, Possler_Daneels_Bowman_2024}, which entail gaining insight and appreciation for the personal search for meaning in life through play \cite{Oliver_Bowman_Woolley_Rogers_Sherrick_Chung_2016, Ryan_Huta_Deci_2008}.

The eudaimonic gaming experience is theorized to have four major components \cite{Possler_2024}: the \textit{motivation}, which encompasses the characteristics of the game and the characteristics of the player; the \textit{game use}, which involves the interaction between the player and the game; the \textit{experience}, which pertains to the experiential and affective elements of the game use; and finally the \textit{effects}, which refer to the outcomes of the experience. Previous research has elaborated extensively on the first three components of this model \cite{Klimmt_Possler_2021, valkenburg2013differential, Cole_Gillies_2021, Cole_Gillies_2022, Rogers_Woolley_Sherrick_Bowman_Oliver_2017, Oliver_Bowman_Woolley_Rogers_Sherrick_Chung_2016}. However, the \textit{effects} of eudaimonic gaming experiences have been relatively underexplored \cite{Daneels_Vandebosch_Walrave_2023}. Current eudaimonic gaming literature focuses on how these experiences may have an effect on the player experience, rather than investigating individual behavioral effects or potential pro-social outcomes \cite{Daneels_Vandebosch_Walrave_2023}. It is highly likely that such meaningful experiences lead to tangible effects; indeed, games have generally been shown to have positive effects on cognitive, motivational, emotional, and social outcomes \cite{franceschini2022short, iacovides2014gaming, greitemeyer2010effects, przybylski2012ideal, granic2014benefits}. As the formation of eudaimonic gaming experiences has been well theorized, there is an opportunity to connect specific elements of its formation to resulting effects. This could improve our understanding of how these experiences shape players' emotions, cognition, and personal growth and allow us to refine existing models of meaningful play. To this end, we ask the following research questions:
\begin{itemize}
    \item \textbf{RQ1}: What perceived outcomes or effects result from eudaimonic gaming experiences?
    \item \textbf{RQ2}: What components of eudaimonic gaming experiences might influence these perceived effects?
\end{itemize}

To answer these questions, we disseminated a survey asking participants (n = 166) to recount a meaningful gaming experience and different aspects of why it may have been meaningful to them. The survey used the extensional definition of eudaimonia \cite{possler2023towards} and was designed around \citet{Possler_2024}'s process model outlining the formation of eudaimonic gaming experiences. We used the submodels \cite{Klimmt_Possler_2021, valkenburg2013differential} and other framing concepts from the process model to develop a deductive scheme for coding the motivation, game use, and experiential elements of eudaimonic gaming experiences. In addition, we inductively coded participants' perceived effects of these experiences and connected our deductive coding scheme to surfaced themes concerning inward-focused, learning, career, social, and health outcomes. Using generalized linear mixed models (GLMMs) to uncover how different aspects of the eudaimonic gaming experience may result in certain effects, we found that personal stress influences perceived health outcomes and narrative insight, while being emotionally moved or challenged influences inward-focused outcomes. We discuss how our findings build upon and extend current models of eudaimonic gaming experiences and identify several open avenues for future theoretical elaboration, as well as opportunities to tie eudaimonic gaming experiences to models of informal learning and joint media engagement. We contribute an empirical mixed-methods understanding of how meaningful experiences with games can lead to long-lasting cognitive, emotional, social, and learning outcomes. 

\section{Related Work}

\subsection{Eudaimonic Experiences}
\label{litreviewsec1}

While the eudaimonic experience is just beginning to be explored in games, its effects have previously been investigated in a large variety of other media formats, such as music \cite{ji2021melody, bhattacharya2023understanding}, movies \cite{tkalvcivc2018eudaimonic, ott2021eudaimonic, rieger2014media}, online videos \cite{janicke2018watching, janicke2021exploring}, and social media interactions \cite{oliver2022social}. Outcomes include altruistic behaviors \cite{schnall2010elevation}, intergroup connectedness, and prejudice reduction \cite{oliver2015media} elicited by inspiring videos, perceptions of self-meaning elicited by music \cite{ji2021melody}, well-being at the workplace elicited by entertaining videos \cite{janicke2021exploring}, and prosocial motivations elicited by commenting and sharing on social media \cite{oliver2022social}. \citet{oliver2021model}'s model of inspiring media specifies three categories of outcomes: affective/cognitive, motivations/behaviors, and general media behaviors.  

Games research, especially in human-computer interaction (HCI), has primarily focused on hedonic experiences \cite{Cole_Gillies_2022, Mekler_Bopp_Tuch_Opwis_2014, Raney_Oliver_Bartsch_2019, Hamari_Keronen_2017}, which encompass pleasure-seeking aspects of the player experience (e.g., flow, presence, and immersion) \cite{Mekler_Bopp_Tuch_Opwis_2014, Mekler_Hornbæk_2016}. However, the last decade has seen rising interest in specifically defining and investigating the eudaimonic gaming experience \cite{Rogers_Woolley_Sherrick_Bowman_Oliver_2017, Mekler_Hornbæk_2016, Possler_2024, Daneels_Bowman_Possler_Mekler_2021, Cole_Gillies_2022, Oliver_Bowman_Woolley_Rogers_Sherrick_Chung_2016, Fröding_Peterson_2013}. \citet{Rogers_Woolley_Sherrick_Bowman_Oliver_2017} found that players primarily defined hedonic or `fun' experiences in terms of gameplay mechanics, while eudaimonic or `meaningful' experiences were defined in terms of social connections with players and identification \cite{Cohen_2017} with in-game characters. Complementary investigations from \citet{Bopp_Mekler_Opwis_2016, Bopp_Müller_Aeschbach_Opwis_Mekler_2019} found that player emotion was often evoked by in-game loss, character attachment, experiencing sadness, and (lack of) agency, which players connected to their personal memories and were able to reflect upon. \citet{Cole_Gillies_2022} developed a grounded theory for emotional exploration to define the eudaimonic experience, elaborate on why players would seek it out, and discuss how developers can design for it. A recent review from \citet{Daneels_Vandebosch_Walrave_2020} found four broad patterns in eudaimonic gaming research: (1) eudaimonic appreciation \cite{Oliver_Bartsch_2010} as an outcome of playing digital games, (2) co-variation among emotionally moving/challenging and self-reflective experiences, (3) eudaimonic social connectedness, and (4) other less-defined eudaimonic concepts such as nostalgia, well-being, and elevation. Building on \citet{Daneels_Vandebosch_Walrave_2020}, \citet{Possler_Bowman_Daneels_2023} developed a definition of the eudaimonic gaming experience: ``Eudaimonic gaming experiences may manifest in various forms, e.g., appreciation, being emotionally moved or challenged, self-reflection, deep social bonds, nostalgia, awe, and elevation.'' Our paper defines the eudaimonic gaming experience as such. In addition, we frame our analysis around \citet{Possler_2024}'s theorized process model, which hypothesizes that eudaimonic gaming experiences are parameterized by \textit{motivation} (seeking meaningfulness as a motive for game use), \textit{game use} (the formation of eudaimonic experiences during playing), \textit{experience} (having a meaningful/eudaimonic experience), and \textit{effects} (outcomes of meaningful/eudaimonic experiences). However, while the first three components of this process model have been well-investigated, the \textit{effects} of eudaimonic experiences are relatively underexplored \cite{Daneels_Vandebosch_Walrave_2023}. This is a gap that this paper seeks to address, as elaborating on the pathways to these effects could inform the design of platforms and interventions targeted at promoting such outcomes. 

\subsection{Factors Affecting the Eudaimonic Gaming Experience} The affordances, mechanics, and design of the game itself are essential to the formation of the eudaimonic gaming experience \cite{Possler_2024}. \citet{Klimmt_Possler_2021} proposed the Agency, Narrative, Sociality, and Aesthetics (ANSA) model, which suggests that the formation of eudaimonic gaming experiences can be explained by the titular four elements. This proposition is well-supported by the literature. \citet{Cole_Gillies_2021}'s framework elaborates on four types of agency potentially relevant to eudaimonic experience: actual, interpretive, fictional, and mechanical.  The mechanical agency afforded by certain game mechanics has been found to be meaningful to players; for example, mastering a set of mechanics to beat difficult challenges in certain games led to a sense of meaningful accomplishment \cite{Rogers_Woolley_Sherrick_Bowman_Oliver_2017}. The fictional agency to make moral and consequential choices in games was also found to be important to players \cite{Iten_Steinemann_Opwis_2018}. Games implement a variety of narrative structures \cite{Carstensdottir_Kleinman_El-Nasr_2019}, and branching narratives in particular allow players to `rehearse their ethos' \cite{Consalvo_Busch_Jong_2019} by actively reflecting upon how their in-game decisions align with their personal values or the personality of the character they are playing as \cite{nay2017meaning}. Indeed, the interpretive agency afforded by narrative can be a powerful tool for transporting players into storyworlds \cite{green2004understanding} and allowing them to identify \cite{Cohen_2017} with characters. Players go so far as to exhibit preferences consistent with those of their player character, role-playing even when not instructed to \cite{Domínguez_Cardona-Rivera_Vance_Roberts_2016, Happ_Melzer_Steffgen_2013}, and sometimes develop deep attachments of a wide emotional range to characters \cite{Bopp_Müller_Aeschbach_Opwis_Mekler_2019}. The sociality of the gaming experience is also a factor of the eudaimonic experience \cite{Daneels_Vandebosch_Walrave_2020, Possler_Kümpel_Unkel_2020}. Games enable multiplayer interactions with both strangers and close others \cite{nardi2006strangers, devasia2025partnership}, as well as 1-on-1 play or turn-taking with siblings \cite{go2012brothers, coyne2016super} and friends \cite{consalvo2018finding}. Finally, the aesthetics—a broad term that refers to the sensory modalities, including visuals, soundtrack, and haptics \cite{possler2023towards}—can prompt intense experiences of appreciation and awe \cite{Bopp_Vornhagen_Mekler_2021}, both key to the eudaimonic experience. 

In addition, player context and characteristics also play a role in eudaimonic gaming experiences, although they have been comparatively less explored \cite{Daneels_Vandebosch_Walrave_2023}. The Differential Susceptibility to Media Effects Model (DSMM) claims that dispositional, developmental, and social variables make up an individual's response to a media object \cite{valkenburg2013differential}. Dispositional factors include sociodemographics, personal values, and affect. A well-researched dispositional motivation is \textit{escapism} (the avoidance of reality to allow for emotional regulation, mood management, recovery, and coping \cite{kosa2020four}) in times of personal crisis \cite{Iacovides_Cox_2015, iacovides2019role}, which was a topic of particular interest over the COVID-19 pandemic \cite{boldi2023making, barr2022playing, turkay2023self}. \citet{stenseng2012activity}'s model of \textit{self-expansive} escapism, in which an individual engages in an activity and views it as an opportunity for personal growth, aligns especially well with the eudaimonic experience. Indeed, some research suggests that players explicitly seek out games that will lead to meaningful personal reflection \cite{Possler_Kümpel_Unkel_2020}, but it has recently been suggested that eudaimonic experiences are coincidentally discovered rather than actively sought after \cite{bowman2024excited}. Developmental variables are relevant during periods of life when a person is undergoing a formative experience \cite{valkenburg2013differential}. Research suggests that eudaimonic gaming experiences are of particular relevance to adolescents; games allow adolescents to self-reflect and practice self-acceptance \cite{Daneels_Vandebosch_Walrave_2020}, and give them an opportunity to `try on' their ideal selves \cite{przybylski2012ideal}. Finally, social variables—which include the presence of friends or other co-players, as discussed above—also play a role in an individual's response to media \cite{valkenburg2013differential}. 

In this work, we use both the ANSA model \cite{Klimmt_Possler_2021} and the DSMM \cite{valkenburg2013differential}, as well as the processes specified by \citet{Possler_2024} surrounding game use and experience, to inform our analysis of the \textit{effects} of the eudaimonic experience. Specifically, we use these models and processes to develop both inductive and deductive coding schemes for our analysis. 

% \subsection{Games and Behavior Change}

\subsection{Self-Determination Theory}
Self-determination theory (SDT) is one of the most common models used to explain the motivational qualities and behavior change evoked by games \cite{Ryan_Rigby_Przybylski_2006, Tyack_Mekler_2020, Tyack_Mekler_2024, tamborini2010defining}. SDT posits that three primary psychological needs form the basis of a human's intrinsic motivation (the human desire for inherently satisfying activities): autonomy, competence, and relatedness \cite{Ryan_Deci_2017}. While autonomy and competence are inherently satisfied by the agency provided by game mechanics and controls \cite{Ryan_Rigby_Przybylski_2006, peng2012need} and are well-researched in game studies, relatedness has been relatively less studied \cite{Tyack_Wyeth_2017, Tyack_Mekler_2024}. \citet{Tyack_Wyeth_2017} point out that the Player Experience of Need Satisfaction \cite{Ryan_Rigby_Przybylski_2006}, a commonly used measure at the intersection of HCI and games, assumes that relatedness is satisfied by multiplayer mechanics. When considering meaningful gaming experiences, this prevailing viewpoint is especially insufficient, as many of the games that elicit eudaimonic responses are single-player \cite{Possler_Daneels_Bowman_2024, Daneels_Malliet_Geerts_Denayer_Walrave_Vandebosch_2021}. \citet{Tyack_Wyeth_2017}'s suggestion that relatedness can also be satisfied through parasocial relationships, games development culture, and the game artifacts themselves merits more exploration in the context of eudaimonic gaming experiences. \citet{Lu_Moller_2024} additionally call for further exploration on how game narratives can fulfill SDT's basic psychological needs, which may also be of particular relevance to meaningful experiences in games. In addition, mastery experiences—closely related to SDT's concept of competence—play a critical role in eudaimonic gaming experiences; \citet{rieger2014media} found that eudaimonic entertainment is strongly associated with mastery experiences, as opposed to the relaxation and detachment linked with hedonic entertainment. Such mastery experiences provide players with internally focused capabilities such as self-efficacy, self-confidence, and resilience, which this work views as possible effects of eudaimonic gaming experiences. 

In a recent review of applications of SDT within HCI and games, \citet{Tyack_Mekler_2024} suggest increasing engagement with certain mini-theories of SDT, rather than models that shallowly engage with its core concepts. Of particular interest to the eudaimonic gaming experience is organismic integration theory (OIT) \cite{pelletier2023organismic}. OIT concerns the internalization and integration of personal values—outcomes that can be fulfilled by game narratives \cite{Lu_Moller_2024}. Parable narratives—stories that track characters' psychosocial development, the most popular of which is the hero's journey \cite{campbell2003hero}—have been suggested to increase the personal relevance of games, which thereby increases internalization of the game's messages \cite{rigby2009virtual}. Barriers to meaningful internalization include parental behavior surrounding play \cite{bradt2024does, van2019parents}, which may be of particular importance to eudaimonic gaming experiences formed during adolescence \cite{Possler_2024}. Additionally, while eudaimonic gaming experiences can theoretically result from stressful life circumstances \cite{iacovides2019role}, the same circumstances can also lead to low need satisfaction, resulting in a predisposition for obsessive play \cite{przybylski2009having}. This work explores the above factors, all of which may affect internalization and integration, within the context of eudaimonic gaming experiences. 

\subsection{Effects of Playing Video Games} While the specific effects of eudaimonic gaming experiences and their formation have not been elaborated upon, the behavioral effects of video game play in general have been well-explored. This is especially true of the \textit{negative} effects of video game play, which remain disproportionately more studied than positive effects \cite{adachi2013video}. In particular, several studies link playing video games with aggressive and violent behavior, e.g., \cite{anderson2001effects, greitemeyer2014video}. However, this is a topic of ongoing debate, e.g., \cite{elson2019policy, ferguson2010much, ferguson2015digital}, and the literature faces continuing concerns surrounding publication bias \cite{hilgard2017overstated, ferguson2015angry}. Hence, we focus this review on prosocial and positive effects of video game play. \citet{granic2014benefits} outlined four main positive effects resulting from video game play: cognitive (e.g., attention, spatial skills), motivational (e.g., resilience), emotional (e.g., self-regulation), and social (e.g., prosocial behaviors). \citet{quwaider2019impact} elaborated on these effect types by specifying player impacts as well as types of games that may lead to certain effects. Cognitive effects, such as visual-spatial skills \cite{green2007action}, selective attention \cite{green2003action}, cognitive enhancement \cite{franceschini2022short}, executive function \cite{eichenbaum2014video}, and problem solving \cite{adachi2013more}, have been especially well elaborated upon. Prosocial and helping behaviors \cite{greitemeyer2010effects}, such as positive humanity traits \cite{greitemeyer2013effects} and increased empathy \cite{greitemeyer2010playing}, can also be promoted by cooperative games and are possibly mediated by identification with characters \cite{Happ_Melzer_Steffgen_2013}. Multiplayer games serve as rich venues for close relationship formation \cite{adachi2013more, verheijen2019influence, kowert2014relationship, martinez2022joint} and maintenance \cite{devasia2025partnership}, with quality possibly mediated by game type \cite{verheijen2019influence} and emotional sensitivity \cite{kowert2014relationship}. While the positive effects of commercial video games on mental health have only been recently explored in the literature \cite{boldi2022commercial, vakeva2025don}, games have proven to have positive effects on emotional regulation \cite{przybylski2012ideal}, identity formation \cite{bassiouni2016video}, mood \cite{hemenover2018video}, affective well-being \cite{johannes2021video}, varying levels of reflection \cite{mekler2018game}, and self-acceptance \cite{johnson2013videogames}, especially for children \cite{kovess2016time}. \citet{halbrook2019and}'s narrative review adds physical health as another benefit of games, especially through exergames such as \textit{Just Dance} \citegame{justdance3} and \textit{Wii Fit} \citegame{wiifit}. \citet{bourgonjon2016players}'s analysis of players' self-reported positive effects also reports on education as an outcome. Indeed, video games have been theorized as a space of informal learning since the advent of game studies \cite{shaffer2005video, shaffer2006computer, gee2003video}. \citet{iacovides2014gaming} proposed the Gaming Involvement and Informal Learning framework, which explores the relationship between identity, involvement, and learning. Informal learning in games can even lead to changes in career \cite{giammarco2015video}. 

In this work, we first intend to compare the effects we uncover as outcomes of eudaimonic gaming experience to pre-existing effects in the literature. Then, we tie elements of the eudaimonic experience to specific effects in order to guide more targeted avenues of exploration. 

\section{Methods}
To investigate what types of perceived effects meaningful experiences in games may lead to, we conducted an exploratory survey touching on the four components of the eudaimonic gaming experience \cite{Possler_2024}. 

\subsection{Participants and Recruitment} The data for this study was collected from October 2024 to January 2025. We collected responses using convenience sampling with a Google Forms survey disseminated through gaming servers and mailing lists at our university and posted on social media sites such as Reddit, Twitter, and Bluesky. We collected a total of 53 initial responses, of which 3 were filtered out due to incomplete or insufficient responses (e.g., left several responses blank, failed the Captcha check). In addition, we posted the survey on the online recruitment platform Prolific. Participants were paid at a rate of 18.72 USD per hour, the recommended Prolific rate at the time of the study. 125 responses were collected, of which 9 were filtered out for similar reasons to above. This resulted in a total of 166 valid responses. All participants were asked to verify that they were over 18 and consent to their responses being anonymously used in this research. The recruitment materials, study protocol, and data collection protocol were reviewed and approved by the Institute Review Board at our university. 

\subsection{Survey} 

\subsubsection{Background Questions}
The survey first asked participants to provide a basic overview of their sociodemographics and gaming behaviors:
\begin{itemize}
    \item Demographics [namely: age, gender identity, race/ethnicity]
    \item What types of games do you normally prefer? [Choose from: Action, MOBA, Puzzle, RPG, Simulation, Shooter, Strategy, Other]
    \item How often do you play games? [six item Likert scale ranging from 1 = 'Never' to 6 = 'Several times a week']
\end{itemize}

\subsubsection{Eudaimonia Questions} \label{eq}
After providing background information, participants were asked to discuss a video game experience that was very meaningful to them in some way and were provided with \citet{Possler_Bowman_Daneels_2023}'s definition to specify what was meant by meaningful.
\begin{itemize}
    \item EQ1: What game was this? [Open-response]
    \item EQ2: What type of game is this? [Choose from: Action, MOBA, Puzzle, RPG, Simulation, Shooter, Strategy, Other]
    \item EQ3: How long ago did you play this game? [Open-response]
    \item EQ4: Can you provide some context as to what your life was like at that time? [Open-response]
    \item EQ5: Why do you think this game experience was so meaningful to you at that time of your life? [Open-response]
    \item EQ6: Is this game still meaningful to your life in any way? If yes, how so? [Open-response]
    \item EQ7: Did this gaming experience cause you to pursue activities that you might not have otherwise? [Open-response]
    \item EQ8: Did this game experience give you a new or different way to see yourself, the way you saw the world, or engaged with others? [Open-response]
    \item EQ9: Have the above changes affected your present life? [Open-response]
    \item EQ10: Do you ever discuss these experiences with others? If so, who? [Open-response]
\end{itemize}

\section{Analysis}
\subsection{Overview} \label{analysisoverview}
We chose to use both deductive and inductive analyses in this paper to ensure alignment of our findings with current research while simultaneously supporting the discovery of new phenomena, particularly for eudaimonic effects of games. We framed our analysis using \citet{Possler_2024}'s process model of the formation of eudaimonic gaming experiences. As associated submodels have been developed for the first three components of the model (motivation \cite{valkenburg2013differential, Klimmt_Possler_2021}, game use \cite{Possler_2024}, and experience \cite{possler2023towards}), we deductively coded data relating to these components to strongly ground our analysis in prior research. The first author developed a codebook around these components, which is presented in the following sections. The research team could add additional “other” codes throughout and were asked to provide specific details on particular codes. However, as the \textit{effects} of eudaimonic gaming experiences are not well elaborated upon in the literature, the questions surrounding effects (EQ7, EQ8, EQ9, and EQ10) were purely inductively coded using the six steps of \citet{Braun_Clarke_2006}'s thematic analysis.  

To ensure consensus on the definitions of the deductive codes, the first, second, and third authors began by applying the codebook to the first 10\% of the data (n = 19 responses).  We used Fleiss' kappa \cite{fleiss1981measurement} or Krippendorff's alpha \cite{krippendorff2011computing} to calculate agreement for our deductive coding scheme. All deductively coded categories that used Fleiss' kappa had moderate to high agreement \cite{landis1977measurement} for the first 10\% of the data and are reported below. As we used Krippendorff's alpha to verify interrater reliability for survey items that were coded with multiple author-defined codes, which previous literature has identified as being inherently difficult to achieve high agreement on \cite{wong2021cross,mcdonald2019reliability}, we accepted any alpha value $\geq$ 0.5 as moderate, similar to ranges provided in \citet{landis1977measurement}. We considered this ``reflect[ive] of a codebook which was meaningful (objective)
but not too restrictive (allowed subjectivity)'' \cite{le2018improving}, and reflective of variations in the amount of context included with items during coding long-form qualitative data with more than one coder \cite{campbell2003hero}. After IRR was calculated and deemed acceptable, the remainder of the data intended for deductive coding was split and coded independently.

For our inductive coding process, we began by independently open coding the first 10\% of the data for questions surrounding eudaimonic effects. We then met to generate an initial set of codes. Subsequently, we performed another round of inductive coding with these codes for the next 10\% of the data; for example, the codes `New skills/knowledge' and `New meaningful connections' were applied to the quote, ``I have a hand-embroidered piece that I made hanging over my work desk and a bunch of new followers on Tumblr/readers on AO3 [Archive of Our Own] now!'' We met for a final time to iteratively discuss, revise, and define codes, and consolidate them into themes using an affinity mapping exercise \cite{scupin1997kj}. See Table \ref{finalcodebook} for the finalized codebook. We then split the entire dataset to recode with finalized themes. These themes constitute our answer to RQ1.

\begin{figure}
  \centering
  \includegraphics[width=\linewidth]{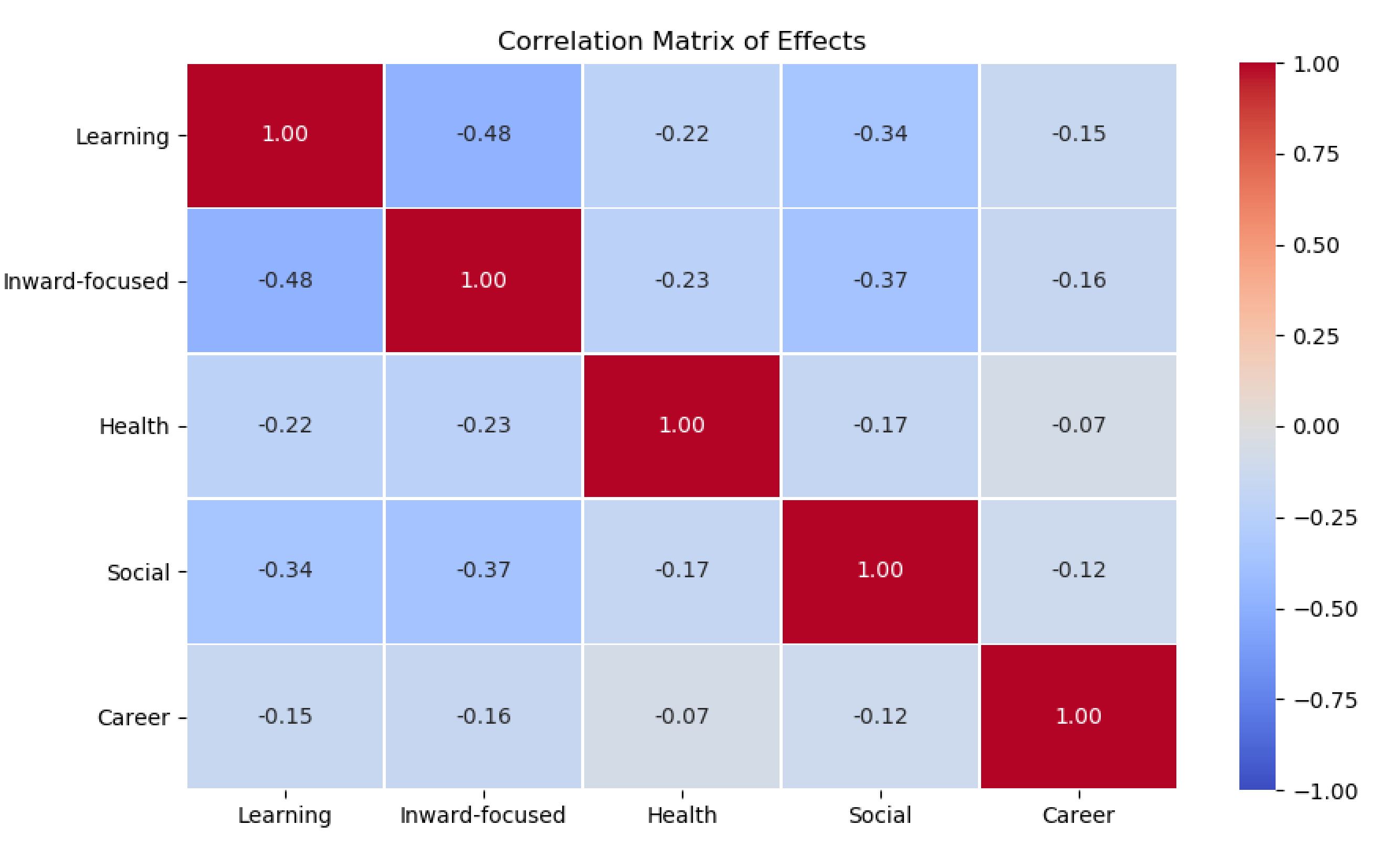}
  \Description{Correlation matrix depicting relations between dependent variables. All correlations are low to moderately weak, implying that outcomes were relatively independent of each other. This justified our approach of modeling all effects separately of each other.}
  \caption{Correlation matrix depicting relations between dependent variables. All correlations are low to moderately weak, implying that outcomes were relatively independent of each other. This justified our approach of modeling all effects separately of each other.}
  \label{matrix}
\end{figure}

To answer RQ2 and examine the relationship between motivation, game use, and experience on our surfaced effects, we used Generalized Linear Mixed Models (GLMM) with binomial logistic regression. We found it appropriate to use binomial rather than multinomial regression, as our dependent variables were not highly correlated with each other (see Figure \ref{matrix}), and we were interested in how distinct sets of predictors—(motivation → effect, game use → effect, and experience → effect)—influenced individual effects (e.g., which motivations may have significantly affected learning). 

After completion of the coding process for RQ1, the authors entered their codes into an Excel spreadsheet, which the first author cleaned with Python. To ensure the data was formatted for the binomial logistic regression model, we one-hot coded Yes/No variables (e.g., Does the game have a clear narrative?) and effect coded categorical variables (e.g., If yes, what type of narrative is it?). In addition, as some variables contained observations with multiple delimited values (e.g., classification of a game's aesthetics), we performed a long-format transformation to the entire data table. Before fitting the model, we dropped variables with extremely high variance inflation factors (VIF), which indicates multicollinearity. We also excluded codes denoted by OTHER from the model, as they did not correspond to well-defined constructs, and chose to focus on modeling concepts pre-defined in the literature. 

\subsection{Coding Motivation}
\citet{Possler_2024} defines motivation for eudaimonic gaming as composed of \textit{game characteristics} (parameterized by the ANSA model \cite{Klimmt_Possler_2021}) and \textit{player characteristics} (parameterized by the DSMM model \cite{valkenburg2013differential}). We use the ANSA model to analyze the game played (EQ1) and the DSMM to analyze player characteristics, detailed in responses to questions EQ2, EQ3, and EQ4. 

\subsubsection{Coding Game Characteristics}
\begin{itemize}
    \item \textit{Agency}: We used \citet{Cole_Gillies_2021}'s framework of agency, which they developed from interviewing participants about their eudaimonic gaming experiences. We ranked games as Low, Medium, or High for each of the four types of agency defined in the paper: Actual Mechanical Agency, which is ``the genuine effect of the players actions and mechanics within the game and analogous to the common understanding of the word agency" ($\kappa = 0.696$), Actual Fictional Agency, which is ``when the player can change the course of the story of the game via their actions, or affect the development and story of other characters in the diegesis'' ($\kappa = 0.504$), Interpretive Mechanical Agency, which ``encourages the player to examine their actions in the game and what they mean when the answers are not made clear to them'' ($\kappa = 0.818$), and Interpretive Fictional Agency, which ``gives the player a minimal narrative framework and encourages them to build their own understanding of the fiction, story and characters'' ($\kappa = 0.894$). 
    \item \textit{Narrative}: We coded the narrative of the reported game using the following questions:
\begin{itemize}
    \item \textit{Does the game have a clear narrative?} We defined narrative as a story that contains event(s), character(s), setting(s), structure, a clear point of view, and a sense of time \cite{Chatman_1978}. We used a Yes/No classification ($\kappa = 0.916$). 
    \item \textit{If yes, what type of narrative is it?} We classified the narrative progression mechanics using \citet{Carstensdottir_Kleinman_El-Nasr_2019}'s framework, which outlines six structure types: \textit{linear}, \textit{branching}, \textit{foldback}, \textit{broom}, \textit{hidden}, and \textit{opportunistic} ($\kappa = 0.864$). 
    \item \textit{If yes, would you consider the narrative to be a parable narrative?} \citet{Lu_Moller_2024} theorize about the possible importance of the \textit{parable narrative} for facilitating personal growth through gameplay. We used their definition to create a Yes/No classification for whether the game narrative could be classified as a parable ($\kappa = 0.779$). 
\end{itemize} 
    \item \textit{Sociality}: We used a binary Yes/No to classify the game by whether or not it had multiplayer mechanics ($\kappa = 0.854$). The sociality of the gameplay itself is addressed in Section \ref{pc:social}. 
    \item \textit{Aesthetics}: Possler characterizes the aesthetic experience as a combination of the visual, auditory, and sensory experiences. We primarily focused on classifying the visual experience using the Video Game Metadata Schema \cite{lee2013developing, VGMS}, a controlled vocabulary for game information that includes visual style as a parameter ($\alpha = 0.751$). As we could not find similar peer-reviewed frameworks for the classification of video game music or haptics, we did not include these in our analysis and acknowledge this as a limitation and potential area for future research. 
\end{itemize}

\subsubsection{Coding Player Characteristics} \label{pc:social}
\begin{itemize}
    \item \textit{Dispositional}: We collected game preferences and sociodemographics from the background portion of the survey. From EQ4, we used a (Yes/No) classification to determine whether the participant was in a time of personal stress or crisis ($\kappa = 0.789$). We supplemented this classification with descriptors from a single round of open coding.  
    \item \textit{Developmental}: We used EQ3 and EQ4 to determine if the player was in a formative period of their life at time of play (ranging from pre-teenaged to early adulthood) with a Yes/No classification ($\kappa = 0.631$). 
    \item \textit{Social}: Using the descriptions provided in EQ4 and EQ5, we determined if the participant played the game 1. alone, 2. with one other person (e.g., partner, sibling), or 3. with multiple people ($\kappa = 0.882$).
\end{itemize}

\subsection{Coding Game Use} \label{gu}
We deductively coded EQ5 using codes derived from \citet{Possler_2024}'s definition of game use ($\alpha$ = 0.590). See Table \ref{tab:game_use_overview} for names and definitions of codes.

\subsection{Coding Experience} 
We developed codes (Table \ref{tab:eudaimonic_definitions}) from the individual components of the extensional definition of eudaimonia \cite{Possler_Bowman_Daneels_2023}. These codes were used to deductively code EQ5 ($\alpha = 0.605$) and EQ6 ($\alpha = 0.664$).
\begin{table}[ht]
\centering
\begin{tabular}{|l|p{8cm}|}
\hline
\textbf{Code Name} & \textbf{Definition} \\
\hline
Appreciation & Broad media gratification as described by \citet{Oliver_Bartsch_2010}. \\ \hline
Sense of meaning & Players connecting game elements to meaningful aspects outside of the game. \\ \hline
Emotionally moved or challenged & Confronting players with emotionally salient material leading to strong mixed-affective responses \cite{bopp2018odd}. \\ \hline
Self-reflection & Players seeking a better understanding of themselves. \\ \hline
Deep social bonds (other players) & Players strengthening relationships with co-players or making new friends through the game. \\ \hline
Deep social bonds (characters) & Players identifying strongly with in-game characters \cite{Cohen_2017}. \\ \hline
Nostalgia & Sentiment or wistfulness for happiness associated with game experience. \\ \hline
Awe & The emotional response to vast and unusual stimuli often involving a sense of amazement and wonder \cite{possler2021awe}. \\ \hline
Elevation & An uplifting and heartwarming feeling as a response to depictions of moral virtue and altruism \cite{Daneels_Vandebosch_Walrave_2020}. \\ \hline
OTHER & Other responses such as escapism, relaxation, etc. \\
\hline
\end{tabular}
\caption{Definitions of codes for eudaimonic experiences during gameplay.}
\label{tab:eudaimonic_definitions}
\end{table}

\subsection{Coding Effects} \label{codingeffects}
EQ7, EQ8, and EQ9 were coded using the multi-round thematic analysis process described in Section \ref{analysisoverview}. The initial open coding process yielded 46 subcodes, which were consolidated into 13 codes, then organized into 5 themes, described in detail in Section \ref{rq1}. EQ10 was coded similarly, with 10 initial codes being narrowed to three main categories, described in Section \ref{effectsoverview}.

\begin{table}[htbp]
\renewcommand{\arraystretch}{0.9}
\small
\centering
\caption{Final version of our codebook, with themes bolded.}
\label{finalcodebook}
\begin{tabular}{|p{6.5cm}|p{6.5cm}|}
\hline
\textbf{1. Inward-Focused} & \textbf{4. Learning} \\
\hspace{1em}a. Acceptance & \hspace{1em}a. New skills/knowledge \\
\hspace{1em}b. Confidence & \hspace{1em}b. Practicing existing skills \\
\hspace{1em}c. Identity & \hspace{1em}c. Individual life skill \\
\hspace{1em}d. Worldview & \\[0.5ex]

\textbf{2. Career} & \textbf{5. Social Connection} \\
\hspace{1em}a. Career-related skills & \hspace{1em}a. People skills \\
\textbf{3. Perceived Health Benefits} & \hspace{1em}b. New meaningful connections \\
\hspace{1em}a. Physical health & \hspace{1em}c. Strengthening existing connection \\
\hspace{1em}b. Mental health & \\
\hline
\end{tabular}
\end{table}

\section{Findings}

\subsection{Descriptive Overview}

\subsubsection{Motivation: ANSA}

In terms of agency, games primarily allowed for high actual mechanical agency, low actual fictional agency, low interpretive mechanical agency, and medium interpretive fictional agency. These attributes were common to single-player narrative RPGs with large open worlds, e.g., \textit{The Legend of Zelda: Breath of the Wild} \citegame{botw}, \textit{Red Dead Redemption 2} \citegame{rdr2}, or \textit{The Elder Scrolls V: Skyrim} \citegame{skyrim}. Indeed, a majority of games (69.2\%) had a clear narrative, and 60.2\% were single-player. Of the narrative games, 52.1\% were classified as having linear or foldback narratives, which have fixed endings. A vast majority of these narrative games (80.2\%) were classified as having a parable framing. The most common visual aesthetics were realistic (39.1\%) or stylized (37.3\%). 

\begin{table}[ht]
    \centering
    \footnotesize
    \begin{tabular}{llr}
        \toprule
        \textbf{Category} & \textbf{Variable} & \textbf{Counts} \\
        \midrule
        \textbf{Agency} & AMA & Low: 30, Medium: 56, High: 80 \\
                         & AFA & Low: 88, Medium: 47, High: 31 \\
                         & IMA & Low: 90, Medium: 54, High: 22 \\
                         & IFA & Low: 52, Medium: 65, High: 49 \\
        \midrule
        \textbf{Narrative} & Has a narrative? & Yes: 115, No: 51 \\
                           & Narrative type & Linear: 44, Branching: 23, Broom: 21, Foldback: 16, Hidden: 11, Not Applicable: 51 \\
                           & Is a parable? & Yes: 93, No: 22, Not Applicable: 51 \\
        \midrule
        \textbf{Social} & Is multiplayer? & No: 100, Yes: 66 \\
        \midrule
        \textbf{Aesthetics} & Visual style & Realistic: 65, Stylized: 62, Map-based: 31, Comic book/Anime: 20 \\
                                   & & Pixel: 16, Cel-shaded: 7, Silhouette: 6, Other: 2, Handicraft: 1 \\
        \bottomrule
    \end{tabular}
    \caption{Counts of each element specified by the ANSA model \cite{Klimmt_Possler_2021} in the experiences reported in our dataset.}
    \label{tab:variables_summary}
\end{table}

\subsubsection{Motivation: Player}
Out of 166 participants, 108 identified as male, 48 as female, 8 as non-binary, and 2 as transgender. Participants ranged in age from 18-61, with an average age of $30.47 \pm 8.37$. Multiple races and ethnicities were represented: White (72.3\%), Asian or Pacific Islander (16.9\%), Black or African American (9\%), and Hispanic or Latinx (9\%). The majority of participants reported playing games several times a week (77.7\%), while the remainder play about once a week (14.4\%), or on a monthly basis (7.8\%). The most commonly preferred genres of games were RPGs (77.7\%), Action (64.4\%), and Shooter (51.8\%). A total of 122 unique games were reported on. The most referenced games were \textit{League of Legends} \citegame{lol} (4.2\%), games from the \textit{Final Fantasy} series (\citegame{ff7, ff8, ff10, ff14, ff15, crisiscore}) (4.2\%), \textit{Disco Elysium} \citegame{discoelysium} (3.6\%), and \textit{Red Dead Redemption 2} \citegame{rdr2} (3.6\%). The reported eudaimonic experiences occurred an average of 7.47 years ago, ranging from within the last month to 30 years prior. While 59.6\% of experiences were within the last 5 years, 23.5\% of players reported that the experience occurred over a decade ago.  

The majority of players (54.2\%) were coded as being in a formative period at the time of the reported experience. These periods included college or graduate school, high school, early adolescence, or other new life contexts such as their first job. While the majority of participants (63.3\%) believed that their life circumstances were relatively normal at the time of play, 36.7\% of participants were coded as undergoing stressful circumstances at the time of play. Examples included social isolation, health or mental health stressors, or abrupt changes in life circumstances. 22.2\% of players explicitly reported that they had started playing the game as a distraction or escape from these circumstances. Most participants played alone in their reported experiences (73.4\%). 18.8\% explicitly mentioned playing with multiple friends, and 7.8\% mentioned playing with one other person, such as a partner or sibling. 

\subsubsection{Game Use}
\begin{table}[ht]
    \centering
    \begin{tabular}{|l|p{8cm}|c|}
        \hline
        \textbf{Code} & \textbf{Description} & \textbf{Count} \\
        \hline
        A/M-BP-1 & Felt a sense of accomplishment & 17 \\
        A/M-BP-2 & Mechanics that surprised or put a twist on game expectations & 6 \\
        N-BP-1   & Gained insight from game narrative & 53 \\
        N-BP-2   & Able to make moral/consequential decisions in the game & 8 \\
        N-BP-3   & Identification with game characters & 20 \\
        S/C-BP   & Played with or against close others & 43 \\
        A-BP-1   & Contemplation of music or graphics & 4 \\
        A-BP-2   & Sense of awe from game interaction & 27 \\
        OTHER    & Something else about the game interaction was mentioned & 35 \\
        Unspecified & Participant did not describe their play experience & 6 \\
        \hline
    \end{tabular}
    \caption{Codes for types of uses the game provides the player in eudaimonic gaming experiences.}
    \label{tab:game_use_overview}
\end{table}

The most common experience that participants described during play of the game itself was gaining insight from the game narrative (31.9\%). This often took the form of self-reflection. For example, P1 stated that \textit{Disco Elysium} \citegame{discoelysium} made them ``consider... [their] values in general, and how [they] want to live life.'' Similarly, play experiences sometimes shifted participants' worldviews. P123 stated that \textit{Red Dead Redemption 2} \citegame{rdr2} ``gave me a new perspective on how valuable and beautiful life is.'' Narrative insight also prompted participants to view other people in a different and more positive light; for example, P125 believed that \textit{Like a Dragon: Infinite Wealth} \citegame{infinitewealth} showed them that ``it is okay to rely on other people and not to take life on by yourself.''

Participants also frequently mentioned the sociality of the play experience, primarily making new social connections through multiplayer games. While these mainly consisted of new friendships, P113 and P117 met their spouses through \textit{Resident Evil 5} \citegame{re5} and \textit{World of Warcraft} \citegame{wow}, respectively. Some participants also deepened social connections with co-players such as partners or siblings. For example, P137 stated about playing \textit{Call of Duty} \citegame{cod} with their partner: ``It’s a time that me and my fiancé get to spend together without the kids, which doesn’t happen often.'' P30 recounted their experience with their siblings: ``Growing up my sisters and I always had to share everything but when it came to the Nancy Drew games...no matter who was playing we would always put our heads to together to figure out the puzzles. We bonded over those games.''

Several aspects of the game use experience were not covered by our coding scheme, the most notable of which was escapism. 19 participants explicitly cited this as the primary benefit of the play experience. For example, P173 stated about their time playing \textit{Skyrim} \citegame{skyrim} during the COVID-19 pandemic: ``I needed to turn my mind off and just live a big adventure and take back, in a sense, the two years I lost inside my house.'' Indeed, of these participants, 13 were coded as undergoing a stressful life circumstance at the time of play. Other factors included novelty or social pressure to play the game, or learning outcomes such as problem-solving skills. 

\subsubsection{Experience} \label{descr_exp}

The most prevalent aspect of the eudaimonic gaming experience that participants felt both at the time of play and in the present was a sense of meaning. This primarily consisted of participants connecting themes and learnings from the game to aspects of their own life. For example, P11 stated that \textit{Chrono Cross} \citegame{chronocross} ``played a significant roles in [their] acceptance of [their] turbulent life experiences instead of lamenting them, while still not wanting them to occur to others.'' P105 said that \textit{Disco Elysium} \citegame {discoelysium} ``completely changed their outlook on life in many regards,'' and P43 elaborated that ``the existentialism of [\textit{Disco Elysium}'s] ending...made [them] feel empowered, in the sense that there is this terrible murder, but there's still a sense of wonder at the world.'' Indeed, of the 44 participants who believed the game gave them a sense of meaning at the time of play, 20 still hold this belief in the present. For the participants who believed the game experience provides a continued sense of meaning, but did not feel so at the time of play, the passage of time likely helped them internalize the game's messages. For example, P167 believes that \textit{To the Moon} \citegame{tothemoon} ``helped [them] understand that happiness in life can come from simple things.'' P119 stated that \textit{Cyberpunk 2077} \citegame{cyberpunk} ``influenced...the values that [they] have today, especially [those] that refer to my mortality and close ones.'' 

While developing a sense of meaning and self-reflecting on the game experience remained fairly constant over the passage of time, other aspects of the eudaimonic experience changed significantly. Being emotionally moved or challenged by the game, as well as feeling strong identification with the characters, seem to be primarily experienced at the time of play. Awe and elevation, which may be tied to the aesthetic elements of the game, also decreased over time. Conversely, nostalgia increased significantly when reflecting on the game experience from the present. 

Although not part of eudaimonia's extensional definition, we included escapism in Table \ref{tab:eudaimonic_counts} due to its frequent appearance in our data. For some participants, the escapism was the most meaningful part of their game experience; e.g., ``it provided a much needed escape from reality where I could actually accomplish things and control my own destiny'' (P12). 

\begin{table}[ht]
\centering
\begin{tabular}{|l|c|c|}
\hline
\textbf{Code Name} & \textbf{EQ5} & \textbf{EQ6} \\
\hline
Appreciation & 24 & 37 \\ \hline
Sense of meaning & 44 & 40 \\ \hline
Emotionally moved or challenged & 28 & 12 \\ \hline
Self-reflection & 25 & 23 \\ \hline
Deep social bonds (other players) & 42 & 28 \\ \hline
Deep social bonds (characters) & 17 & 6 \\ \hline
Nostalgia & 2 & 36 \\ \hline% Combined nostalgia + nostagia
Awe & 16 & 8 \\ \hline
Elevation & 9 & 1 \\ \hline
Other (escapism) & 25 & 8 \\ \hline
OTHER & 12 & 21 \\  \hline% Combined all "other" categories
Not meaningful & 0 & 12 \\
\hline
\end{tabular}
\caption{How affective elements of the eudaimonic gaming experience affected players both at the time of play (EQ5) and in the present (EQ6).}
\label{tab:eudaimonic_counts}
\end{table}

\subsubsection{Effects} \label{effectsoverview} 78.3\% of participants believed that the recounted experience has an effect on their current lives. The types of effects are summarized in Table \ref{tab:effects_overview} and elaborated on in Section \ref{rq1}. When asked if they had discussed these experiences with others, 34.3\% said that they had not, with some participants citing embarrassment or perceived judgment as a reason. 10.2\% of participants said that they shared their experiences with their spouses or partners, and 17.5\% reported sharing with their family members, such as siblings or cousins. Notably, only one participant explicitly reported sharing with their parents. Sharing with social circles of close friends (39.2\%) or peers (11.4\%), such as coworkers or other players of the referenced game, was also common. A small percentage (3\%) of participants stated that they were very open about their experiences and were willing to share with anybody, including strangers.

\begin{table}[ht]
    \centering
    \begin{tabular}{|l|p{8cm}|c|}
        \hline
        \textbf{Effect} & \textbf{Definition} & \textbf{Count} \\
        \hline
        Inward-focused & The game made participants' perception of themselves and the world more positive. & 63 \\ \hline
        Learning & The game inspired participants to engage in skills-based activities. & 58 \\ \hline
        Social connection   & Participants strengthened or formed new relationships with close others. & 39 \\ \hline
        Perceived health benefits   & The game positively affected participants' physical and/or mental health. & 18 \\ \hline
        Career   & The game affected participants' career choices. & 9 \\ \hline
        Unspecified  & Participants believed the game has an effect on their current lives, but did not specify how. & 19 \\ \hline
        Not applicable  & Participants did not believe the game affected their current lives. & 21 \\
        \hline
    \end{tabular}
    \caption{Developed themes for effects of eudaimonic gaming experiences and their definitions (see Section \ref{rq1} for further elaboration).}
    \label{tab:effects_overview}
\end{table}

\subsection{Perceived Effects of Eudaimonic Gaming Experiences} \label{rq1}
Through our inductive coding process for effects, constituting our answer to RQ1, we developed the following five themes: 
\subsubsection{Inward-focused} \label{rq1if} The most frequently reported effects (38\%) were inwardly directed changes that involved changes in personal worldviews, increases in self-confidence, identity formation, and self-acceptance. These outcomes were likely formed through self-reflection prompted by the eudaimonic gaming experience. Participants discussed how their experiences changed the way they viewed society and the world around them; e.g., ``[\textit{Halo: Combat Evolved}] \citegame{haloce} made [P71] see the world more as a place to explore and experience,'' or \textit{Outer Wilds} \citegame{outerwilds} made P165 ``more protective and appreciative of those around [them].'' These worldview shifts were not always positive; for example, P112 believed that \textit{Metal Gear Solid} \citegame{mgs} made them realize the world was ``a harsh environment where everything is based on how much money somebody has.'' However, as the majority of games contained parable narratives, the depictions of protagonists battling and winning against dire circumstances seemed to inspire participants to the same mindset overall; e.g., even though \textit{Red Dead Redemption 2} \citegame{rdr2} made P163 ``see how toxic our current relationship with the world is,'' they believed that the game ``helped [them] see...[that] even if the situation is bad, to make the best of it.'' Shifts in personal worldviews also encompassed how participants viewed society and the people around them. For example, \textit{The Walking Dead} \citegame{thewalkingdeaddefinitive} ``taught [P29] how much strangers can be kind to each other in times of need, but also how inhumane they can be.'' \textit{The Last of Us} \citegame{lastofus}, another game about an apocalyptic pandemic ravaging the world, ``made [P65] see the world, the [COVID-19] pandemic, and the situations that were occurring around [them] in a different way and allowed [them] to relate more.'' Worldview shifts sometimes had broader pro-social outcomes; e.g., P144 discussed how \textit{Animal Crossing: New Horizons} \citegame{acnh} ``helped [them] learn how giving back to others is a wonderful thing to do'' and now volunteers every two weeks as a result.

Participants discussed how their gaming experiences helped them develop self-confidence and self-esteem. This prompted participants to be more open to new experiences; e.g., P34 believed that \textit{Outer Wilds} \citegame{outerwilds} made them ``more willing to step out of [their] comfort zone and try things,'' or P73's statement that ``knowing what [they were] capable of [in \textit{Factorio} \citegame{factorio}] gave [them] confidence to try learn a new skill.'' Many participants believed that their experiences also helped them develop self-esteem; e.g., P173 discussing how \textit{Dragon Age: Inquisition} \citegame{dainquisition} made them ``decide to pursue [graduate] studies thanks to [their] increased self-esteem,'' or \textit{Persona 4 Golden} \citegame{p4g} helping P128 ``feel a lot more confident [and] not feel threatened by people who call me a loner.'' Participants also developed confidence around social interactions, even through single-player games; e.g., \textit{The Legend of Zelda: Link's Awakening} \citegame{linksawakening} ``helped [P48] be more confident in engaging with strangers when needed''; P52 believes that \textit{Skyrim} \citegame{skyrim} enabled them to ``be more bold in a lot of social interactions and...just speak my mind without being terrified of what the other person might say.'' 

Eudaimonic gaming experiences also helped participants form their identities and develop a sense of self-acceptance. These changes likely ``formed through introspection'' (P11, \textit{Chrono Cross} \citegame{chronocross}) and self-reflection. Indeed, games likely gave players a sense of ``playfulness with [their] identities, [allowing them to] be whoever [they] want to be'' (P43, \textit{Disco Elysium} \citegame{discoelysium}). For example, \textit{Shadow of the Colossus} \citegame{shadowofthecolossus} ``made [P176] reflect on [their] own values and what [they] really want to stand up for...remind[ing] [them] that life doesn't have easy answers...and to think about the effects of [their] actions, not just what [they] hope to achieve.'' Positive changes in identity formation sometimes bestowed participants with powerful protective elements that helped them through difficult situations in their lives. P157 ``recently went through the worst phase of their life'', but continued to think about what the protagonist of \textit{Crisis Core: Final Fantasy VII} \citegame{crisiscore} would do in the same situation. The experience ``changed [their] way of seeing life and how [they] treat others [by] trying to be the best version of [themselves] by being kind and charitable.'' Similarly, \textit{Final Fantasy X} \citegame{ff10} helped P118 ``see a light at the end of the tunnel in the form of being able to better myself and grow in my relationships.'' 

% IDENTITY AND SELF ACCEPTANCE

% "something about playfulness with my own identity… being whoever i want to be, and being open to the world's wonders" 

% " find a new way to find joy in my body" P51 Just Dance

% "more importantly gave me more acceptance for the differences I have." Life is Strange

% "It is still meaningful to my life currently because It has helped me grow as a person. 

\subsubsection{Learning} \label{rq1learning} A large percentage of participants (34.9\%) believed that the eudaimonic gaming experience had positive effects on their learning. Often, this entailed informal learning about various framing elements of the game narrative, such as philosophy and politics (P1, \textit{Disco Elysium} \citegame{discoelysium}), mythology (P90 and P107, \textit{Age of Mythology} \citegame{aom}), finance and the economy (P69, \textit{Runescape} \citegame{runescape}), and history (P92, \textit{Indiana Jones and the Great Circle} \citegame{indygreatcircle}). Other participants started new hobbies related to the game, such as archery (P173, \textit{Skyrim} \citegame{skyrim}), karate (P158, \textit{Ghost of Tsushima} \citegame{ghostoftsushima}), or kickboxing (P93, \textit{Tekken} series \citegame{tekken}). Participants also learned to recreate aesthetic elements of the game, such as P151 learning guitar to play songs from \textit{The Last of Us} \citegame{lastofus}, P176 creating background score music that ``reminded [them] of \textit{Shadow of the Colossus} \citegame{shadowofthecolossus}'', P8 learning how to set puzzles to emulate the in-game puzzles from \textit{Ace Attorney} \citegame{aa}, P171 learning 3D modeling to recreate the ship from \textit{Outer Wilds} \citegame{outerwilds}, or P20 and P157 creating fanfiction and fanart of the characters from \textit{Undertale} \citegame{undertale} and \textit{Crisis Core: Final Fantasy VII} \citegame{crisiscore} respectively. Some participants were even inspired to develop their own games (P35, P39, P73, P96, P120) or to learn a language to more deeply engage with the game and/or fellow players (P58, \textit{League of Legends} \citegame{lol}; P121, \textit{Final Fantasy XIV} \citegame{ff14}). P159 elaborated upon why they—and perhaps the aforementioned participants—believed that their interest in games made them more interested in learning: ``Studying in school is boring and doesn't contribute much to life...but studying [about] \textit{Caesar III} \citegame{caesar3} is interesting and [allowed them to] passionately explore the world in different ways.'' In fact, P146 credited their eudaimonic gaming experience with completing school at all, stating that ``[they are] neurodivergent, so being able to get through high school and college was a miracle for [them]...without games [they] don't know if [they] would have learned or cared to persevere in anything.'' 

Participants also believed that they learned valuable life skills from their experiences in games. Some accredited games with teaching them fundamental skills, such as cooking (P133, \textit{Final Fantasy XV} \citegame{ff15}) or driving (P88, \textit{Grand Theft Auto: San Andreas} \citegame{gtasanandreas}). Cognitive skills included problem solving (P56, \textit{League of Legends} \citegame{lol}), critical thinking (P61, \textit{Sparta: War of Empires} \citegame{sparta}, and multitasking (P168, \textit{Tetris} \citegame{tetris}). In terms of motivational benefits, P143 and P165 learned the importance of persistence, stating that ``\textit{CS:GO} \citegame{csgo} taught [P143] that if [they] want to achieve something, everything is possible with hard work" and that "\textit{osu!} \citegame{osu} showed [P165] how satisfying it can be to improve at something.'' Notably, both games are mechanically difficult to master. Similarly, \textit{Dark Souls} \citegame{darksouls} ``taught [P67] to never give up, even when faced with impossible odds.'' P164 elaborated on how the persistence they developed playing \textit{Need for Speed: Most Wanted} \citegame{nfs} shapes their current life, saying that ``staying focused and learning from setbacks has become a guiding principle in [their] life; when faced with obstacles now, [they are] more likely to view them as opportunities to refine [their] skills or develop new strategies rather than feeling defeated.'' 

\subsubsection{Social connection} \label{rq1sc} Eudaimonic gaming experiences helped participants form new meaningful connections that shaped their lives as they are today. For example, P32 believed that ``the connections that [they] built through [\textit{Minecraft}]...probably made me more well adjusted today.'' P95, who also spoke about \textit{Minecraft} \citegame{minecraft}, summarized: ``My current life is a result of my friendships.'' Some participants met their partners through games, such as P7, who credited \textit{World of Warcraft} \citegame{wow} for ``mov[ing] to a different state and start[ing] a new relationship, [getting] married, and [having] a child—all things [they] did not plan on doing before.'' P117 also stated that they are currently married to someone they met in \textit{World of Warcraft}. While not a multiplayer game, P167 said that \textit{To the Moon} \citegame{tothemoon}—a game about a dying man's scattered memories of his deceased wife—``helped [them] `seize the moment' with [their] now-husband and daring to start a relationship with him,'' saying that they were ``not sure if [they] would have done it without [\textit{To the Moon}], because [they were] quite timid...and living behind a shell.'' In addition to new relationships, meaningful connections included joining game-related online communities, and engagement in these spaces often outlasted the original game experience by several years. P8 and P20 referenced ongoing participation in the Phoenix Wright \citegame{aa} fanfiction and Undertale \citegame{undertale} fanart communities, respectively. P86 learned to mod \textit{Baldur's Gate 3} \citegame{bg3} by ``seeking help on Discord [servers]'' related to the game. 

Meaningful gaming experiences also helped participants strengthen existing relationships with close others. These included bonding experiences with partners, family, and friends. P38 uses playing \textit{Mario Kart 64} \citegame{mariokart64} with their wife as a time to ``show her how much I love her,'' even when being competitive. \textit{It Takes Two} helped P31 realize that ``the mundane every day chores...were a lot more fun when done with a partner.'' P78 believed that \textit{Spiritfarer} \citegame{spiritfarer}, a game themed heavily around loss and grief, ``made [them] feel closer to their family and friends [and] made [them] appreciate them even more.'' P118 felt similarly about \textit{Final Fantasy X} \citegame{ff10}, a game with similar themes, and said that ``the lessons [they] learned from it have definitely stuck...and [their] relationships with those around [them] are stronger as a result of those lessons.''  

Participants also believed that ``interacting with other players gave [them] a different perspective on the world at large'' and ``made [them] more open-minded'' (P121, \textit{Final Fantasy XIV} \citegame{ff14}). In particular, these gaming experiences helped them develop valuable people skills, such as collaboration and widening cultural mindsets. For example, P4 believed that \textit{Dofus} \citegame{dofus}, an MMORPG, helped them ``collaborate with people from different countries with different culture.'' Similarly, P30 stated that playing \textit{Nancy Drew and the Curse of Blackmoor Manor} \citegame{nancy} with their siblings ``set up the foundation [for] how [they] interact and work in group settings.'' P22 ``developed a broader mindset due to [the] cultural representations'' portrayed in \textit{GTA San Andreas} \citegame{gtasanandreas} by "reading quite a bit about the region the game was set in.'' Participants were also able to ``practice social skills'' (P10, \textit{The Legend of Zelda: Ocarina of Time} \citegame{oot}, such as ``how to be more confident when interacting with strangers'' (P37, \textit{Duet Night Abyss (beta)} \citegame{dna}), and overall be ``a little more courageous in trying new things and meeting new people'' (P101, \textit{Metaphor Re:Fantazio} \citegame{metaphor}). 

\subsubsection{Perceived health benefits} \label{rq1health}
10.8\% of participants believed that the eudaimonic gaming experience had positive effects on their physical and/or mental health. Some participants were motivated to engage in physical activities due to wishful identification with game characters; for example, P22 believed that \textit{GTA San Andreas} \citegame{gtasanandreas} ``inspired [them] to learn swimming after [COVID-19] lockdown because one of the features was that your character could swim when you jump in the water.'' Other participants were inspired by the naturalistic framing of the game itself; for example, P38 said that \textit{Rain World} \citegame{rainworld} ``had some influence in my decision to learn to meditate and to spend more time hiking and in nature.'' Increasing physical health can correspond to increases in mental health, as pointed out by P17, who believed that \textit{A Short Hike} \citegame{ashorthike} motivated them ``to do more things outside like [they] did before the pandemic, and that had a positive impact on [their] mental health.'' The mechanical affordances of certain games also promoted physical activity. P51 recalled that \textit{Just Dance 3} \citegame{justdance3} ``helped [them]...take care of my health, which improved my self-image and mental well-being over time.'' Likewise, P140 said that \textit{Batman: Arkham Knight} \citegame{arkhamknight} was ``the first game that allowed [them] to move [their] body in a meaningful way while playing it.'' 

In terms of mental health benefits, game experiences were credited with strengthening emotional regulation, from learning mindfulness (P106, \textit{League of Legends} \citegame{lol}) and patience (P108, \textit{Elden Ring} \citegame{eldenring}), to healthy coping mechanisms; e.g., P78's belief that \textit{Spiritfarer} \citegame{spiritfarer} ``helped [them] manage [their] emotions about the ones [they] lost and helped [them] cope in healthy ways.'' Similarly, P169 said that \textit{Persona 3} \citegame{persona3} ``helped [them] grieve and overcome loss much more easily in [their] life.'' Instances of strong identification with the game narrative and characters also had positive effects on participants' mental health. For example, P24, who now works in games and mental health, stated that \textit{Final Fantasy VII} \citegame{ff7} showed them how they ``could be part of other's journeys toward processing, healing, and connecting in the world.'' P9 believed that their experience playing \textit{Dragon Age: Inquisition} \citegame{dainquisition} enabled them to ``find a sense of belonging within [themselves] and [that they] can find it without needing another person to make them feel it.'' P27 gave a powerful account of the resilience they developed through identifying with a character from the \textit{Danganronpa} series: ``One of the survivors of the first game \citegame{danganronpa1} is a severely mentally ill girl who I saw a lot of myself in at the time, and a later game in the series shows her still ill, not at all cured, but with a newfound determination. I started earnestly taking steps to treat my depression and not succumb to despair. Where I used to take medicine and go to therapy only because my family told me to, I was now doing it for myself.'' Increased sociality also proffered benefits to participants' mental health, e.g., P7 stating that \textit{World of Warcraft} \citegame{wow} ``helped [them] come out of long term isolation and make friends and have hope again.'' 

\subsubsection{Career} \label{rq1career}
A small percentage (5.4\%) of participants believed that the eudaimonic gaming experience affected their careers. This primarily occurred through pursuing skills that participants built around the game and related interests. For example, P20 believes that creating \textit{Undertale} fanart started their career as a professional illustrator. P29, in their final year of university, stated, ``I never thought I'd want to consider a career in games, but because of how games have impacted me since childhood that's changed.'' P42 went as far as to say that ``[they] used to consume so much \textit{Minecraft} \citegame{minecraft} content, [they] also almost interned there just because of how much the game meant to [them] and is probably why [they] work at Microsoft,'' adding that they gained ``HR and management skills through moderating...[due to] having a lot of responsibility in the [\textit{Minecraft}] server.'' P4 and P6's interest in how games were built (P4) and experience hosting websites for game content (P6) led them to careers in software engineering. P67 described a more affective path from the game to their career, saying that \textit{Dark Souls} \citegame{darksouls} ``was one of the main elements that influenced [them] to get into one of the hardest schools of law in [country of residence], something that [they] never thought [they] would have been capable of doing.'' 

\subsection{Subcomponents of Perceived Effects} \label{rq2} 
In answering RQ2, we first determined the most prominent subelements of each effect, which are displayed in the relevant diagrams. Our expectation was that the model would likely find these subelements to be statistically significant and positively correlated (denoted by green highlighting in the diagrams) to the formation of the effect in question. We found that while our expectations partially aligned in this regard, not all subelements that were high in frequency corresponded to significant influence on the effect. Conversely, subelements that were particularly low in frequency per effect were sometimes negatively correlated (denoted by red highlighting in the diagrams) to its formation. 
\subsubsection{Inward-focused} 
\begin{figure}
  \centering
  \includegraphics[width=\linewidth]{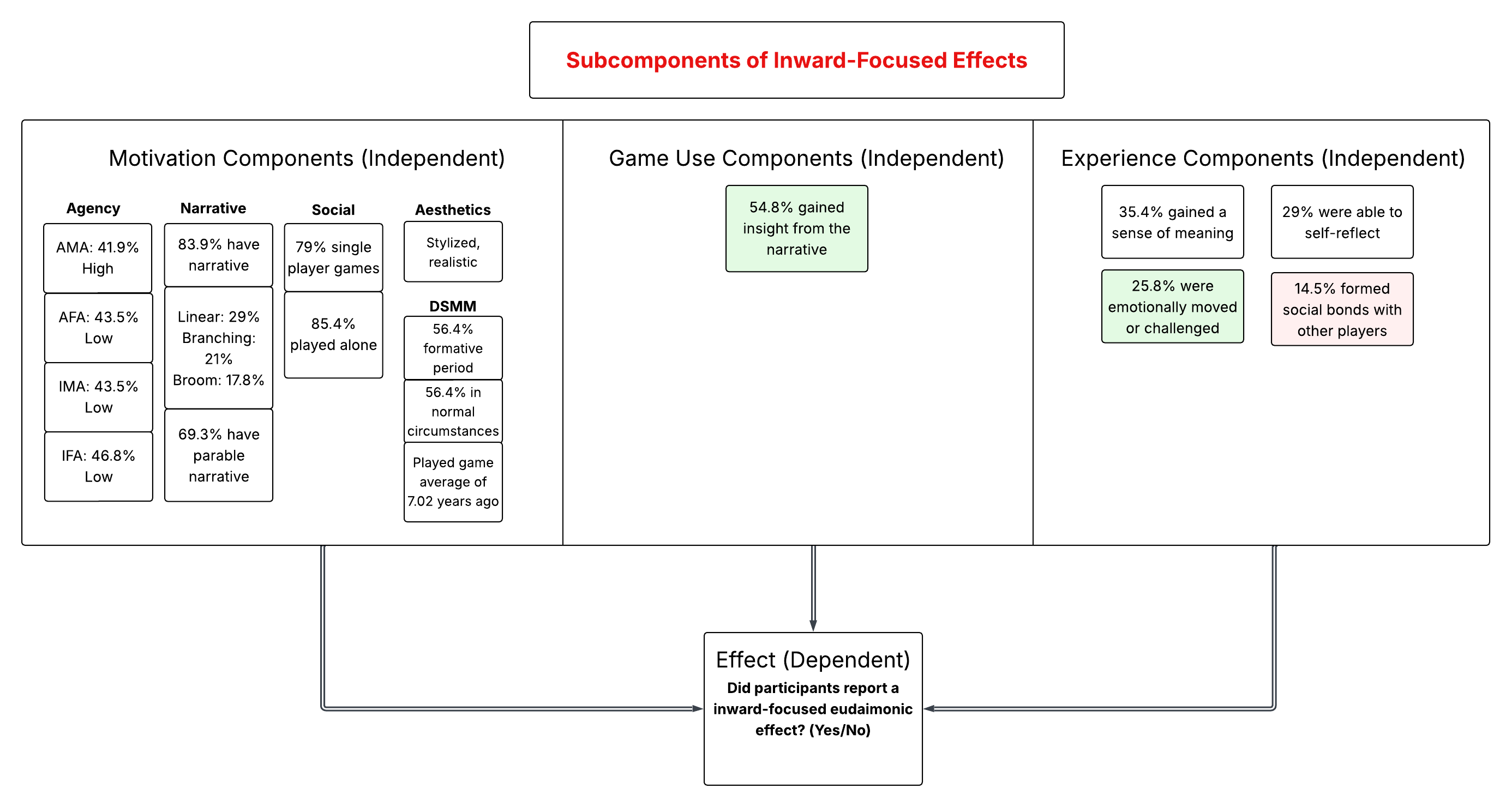}
  \Description{Results from the three binomial logistic regression models (motivation/game use/experience -> formation of an inward-focused effect.}
  \caption{Results from the three binomial logistic regression models (motivation/game use/experience -> formation of an inward-focused effect).}
  \label{motiv_results}
\end{figure}

A total of 62 game experiences were reported to lead to inward-focused outcomes. While the motivational model revealed no significant elements (see Figure \ref{motiv_results}), the most common (54.8\%) game use surrounding gaining insight from the narrative was significant and positively correlated in our game use model ($\beta$ = 0.94, p = 0.013). Relatedly, the experience of being emotionally moved or challenged was significant and positively correlated ($\beta$ = 0.197, p = 0.049) in the experience model. Bonds with other players was a significant but negatively correlated factor in the experience model ($\beta$ = -0.167, p = 0.041), likely because social experiences were not frequently mentioned in the subset of data focused on inward-focused outcomes (14.5\%). 

\begin{figure}
  \centering
  \includegraphics[width=\linewidth]{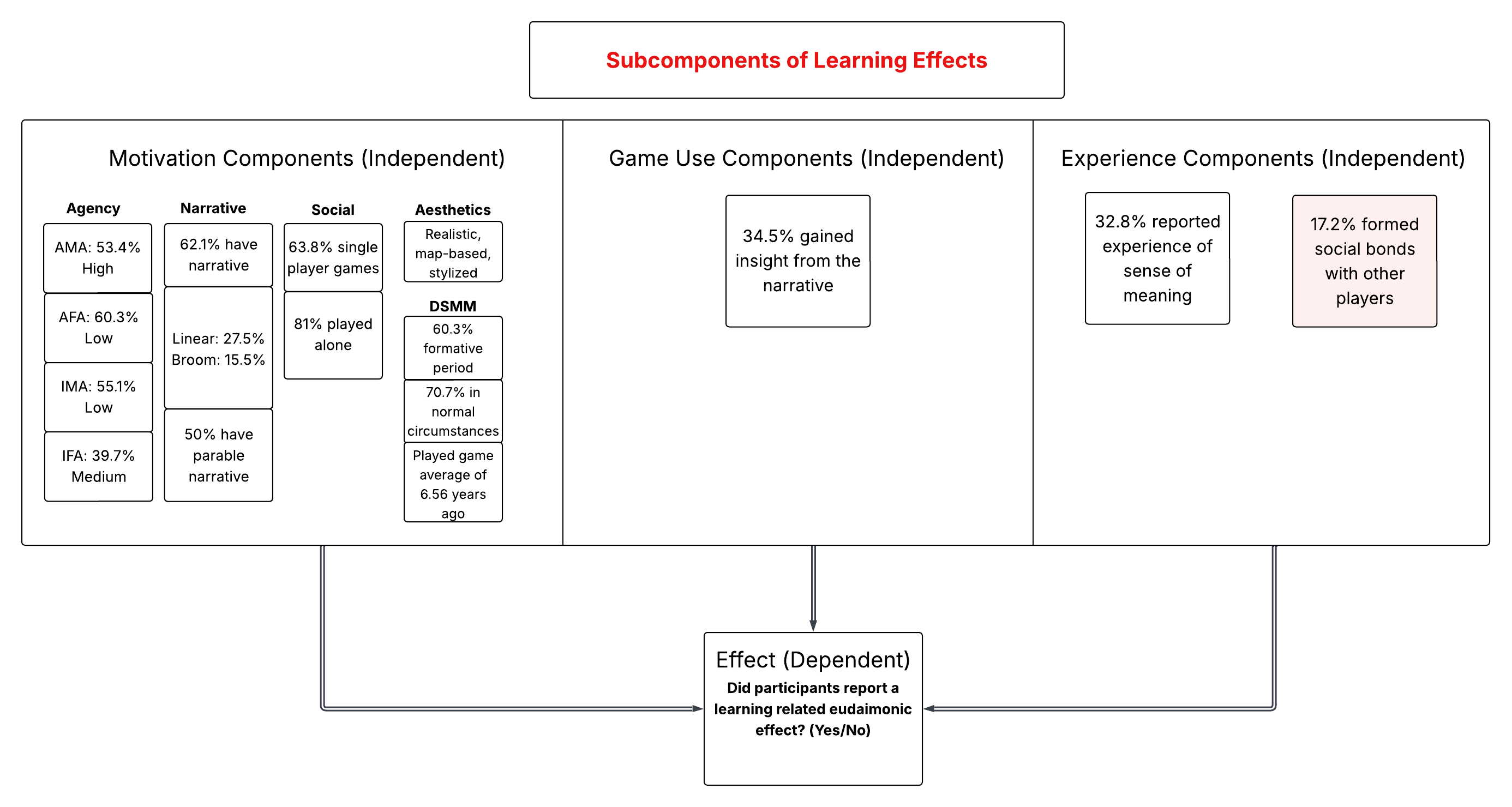}
  \Description{Results from the three binomial logistic regression models (motivation/game use/experience -> formation of a learning related effect).}
  \caption{Results from the three binomial logistic regression models (motivation/game use/experience -> formation of a learning-related effect).}
  \label{learning_results}
\end{figure}
\begin{figure}
  \centering
  \includegraphics[width=\linewidth]{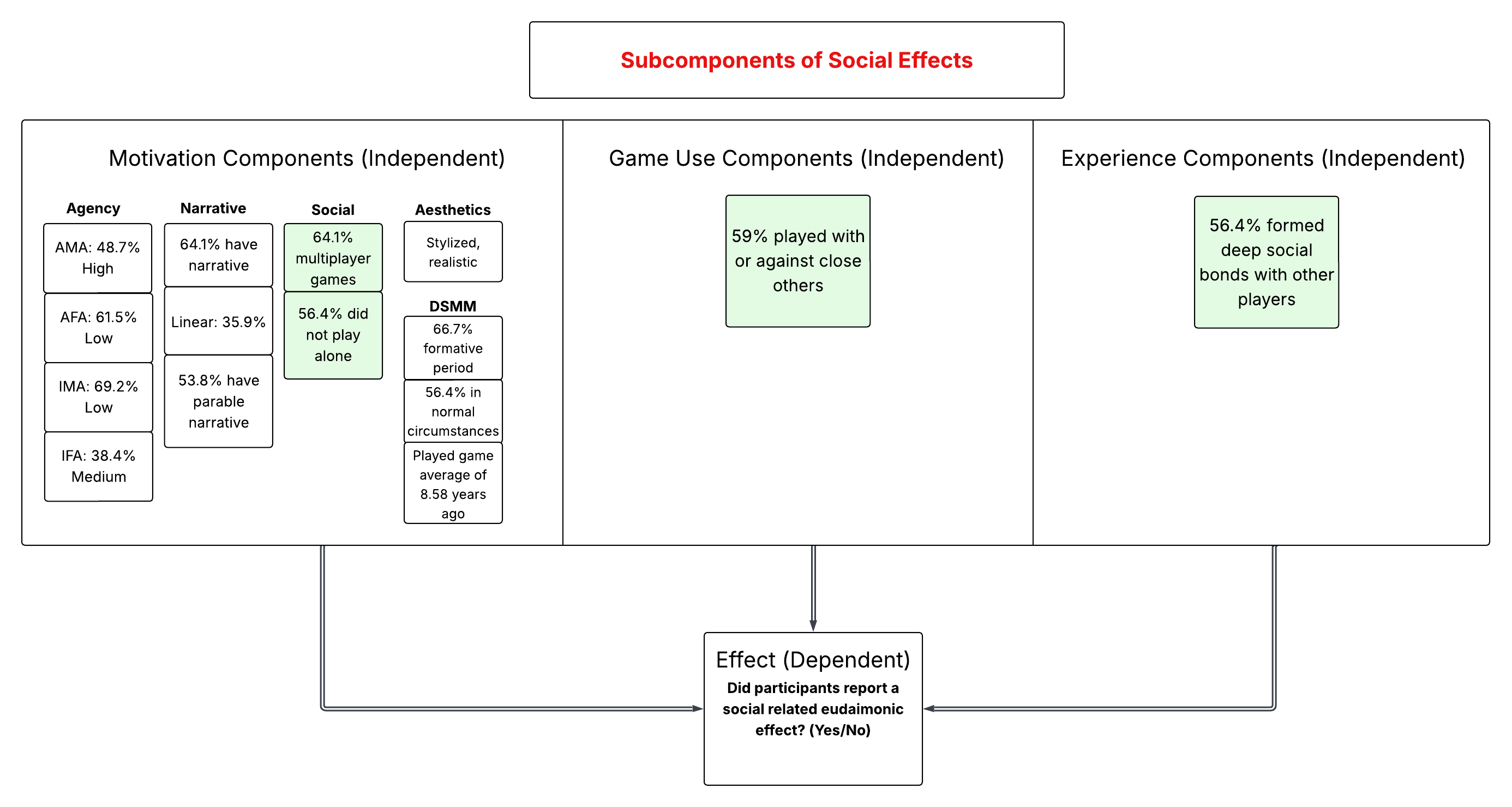}
  \Description{Results from the three binomial logistic regression models (motivation/game use/experience -> formation of a social related effect).}
  \caption{Results from the three binomial logistic regression models (motivation/game use/experience -> formation of a social-related effect).}
  \label{social_results}
\end{figure}
\begin{figure}
  \centering
  \includegraphics[width=\linewidth]{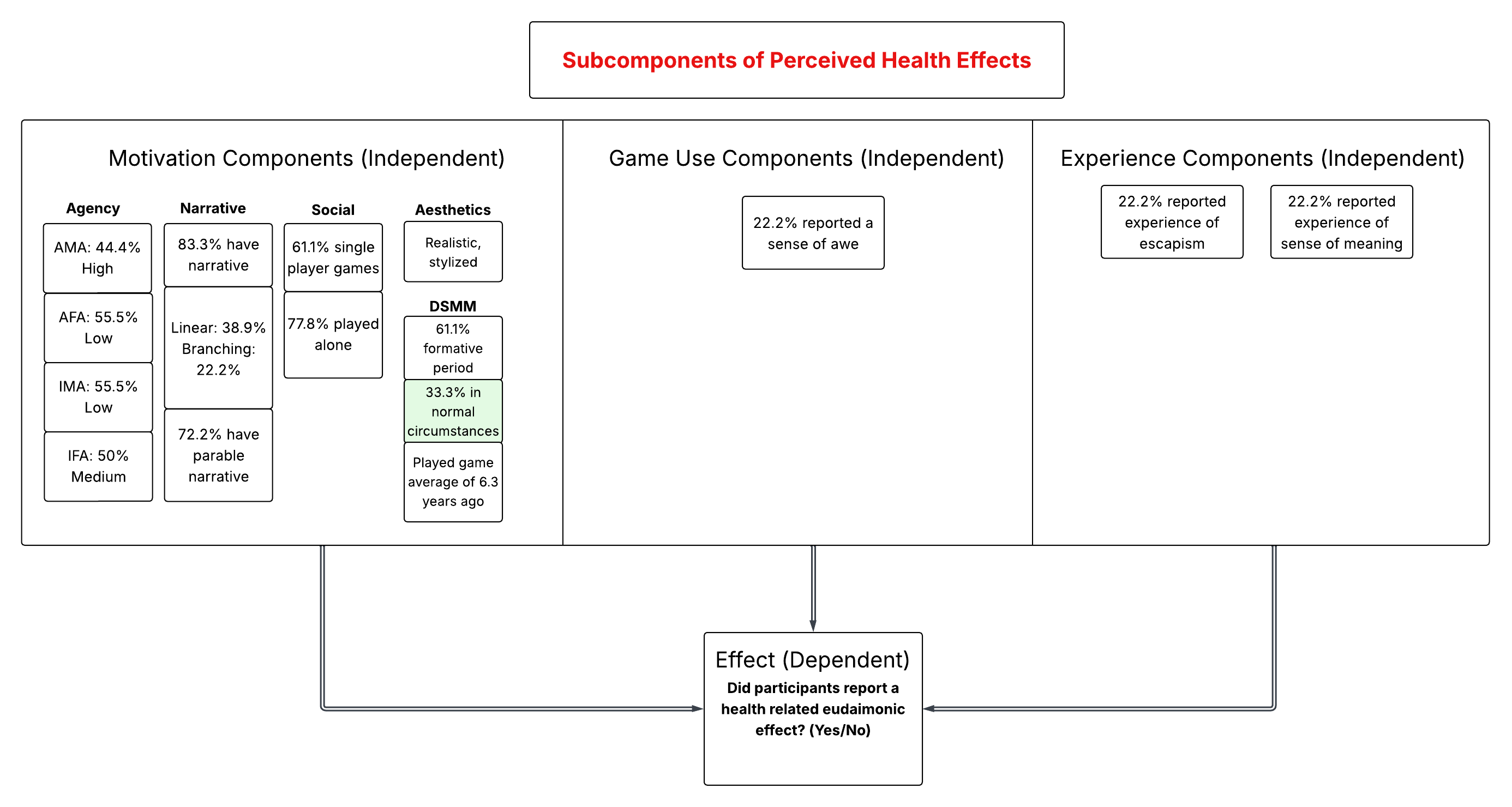}
  \Description{Results from the three binomial logistic regression models (motivation/game use/experience -> formation of a perceived health-related effect).}
  \caption{Results from the three binomial logistic regression models (motivation/game use/experience -> formation of a perceived health-related effect).}
  \label{health_results}
\end{figure}
\begin{figure}
  \centering
  \includegraphics[width=\linewidth]{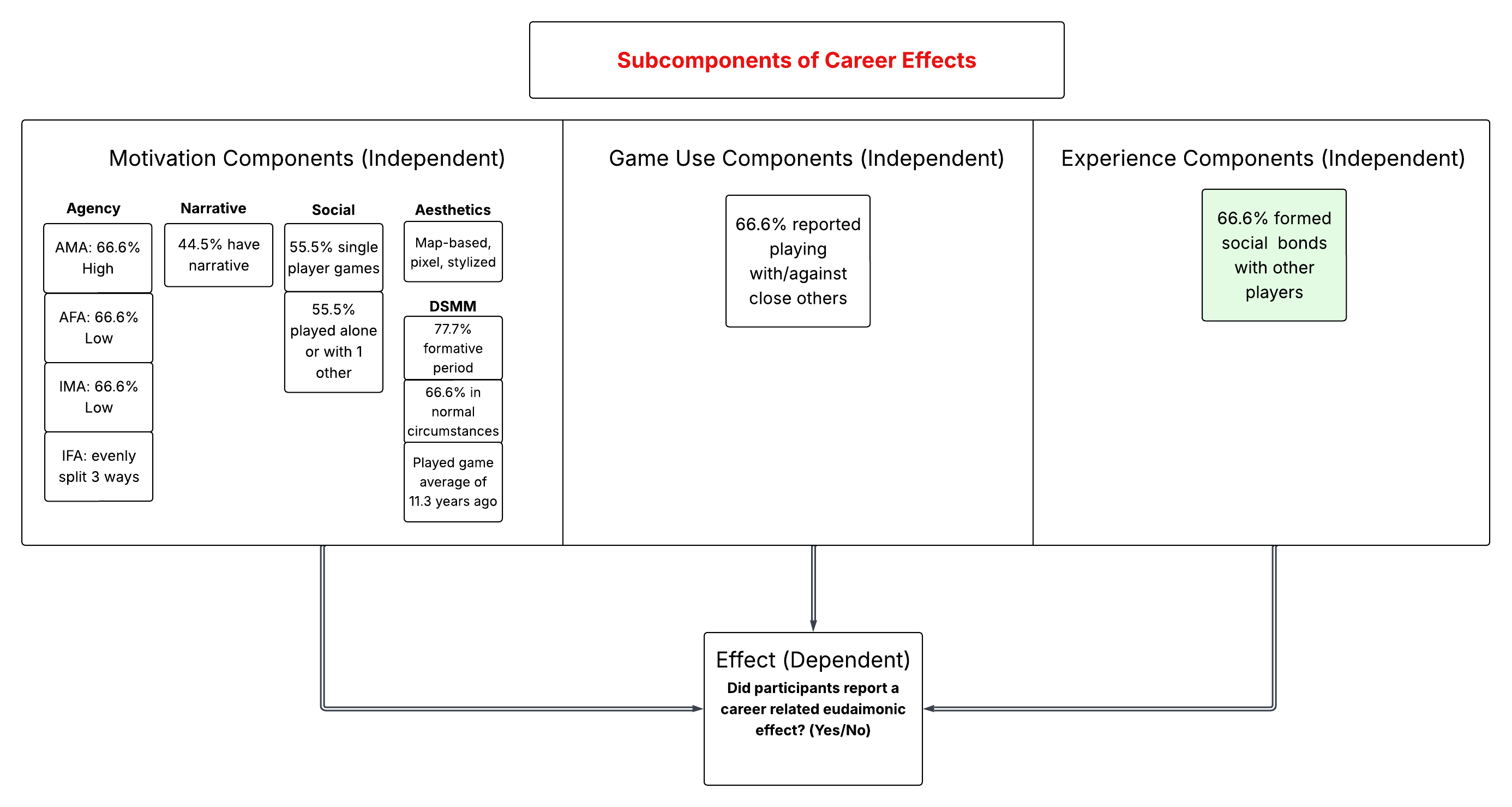}
  \Description{Results from the three binomial logistic regression models (motivation/game use/experience -> formation of a career-related effect).}
  \caption{Results from the three binomial logistic regression models (motivation/game use/experience -> formation of a career-related effect).}
  \label{career_results}
\end{figure}

\subsubsection{Learning} 

A total of 58 game experiences were reported to lead to learning outcomes. We did not observe any of the predominant motivations, game uses, or experiences having an effect on learning (see Figure \ref{learning_results}); however, bonds with other players, a relatively uncommon experience in this category (17.2\%) was a significant but negatively correlated experience ($\beta$ = -0.167, p = 0.044). This may suggest that learning-related effects primarily form through individual play. 

\subsubsection{Social connection} 

A total of 39 game experiences were reported to lead to positive social outcomes. Figure \ref{social_results} shows that any motivations, game uses, or experiences related to the social experience of the game correlated to a social effect. Unsurprisingly, the game being multiplayer (64.1\%) was significant and positively correlated ($\beta$ = 1.38, p = 0.009), as well as the sociality (56.4\% played with multiple people or one other person) of the play experience ($\beta$ = 0.732, p = 0.013). The majority of participants played with or against close others (59\%). Likewise, the most common experience was deep social bonds with other players (56.4\%), and both factors were significant and positively correlated in their respective models ($\beta$ = 1.763, p < 0.001; $\beta$ = 0.308, p < 0.001, respectively).

\subsubsection{Perceived health benefits} \label{rq2health} 

A total of 18 game experiences were reported to lead to health benefits. 50\% of participants were female and 22.2\% were non-binary; although not included in our model, we observed this to be notably different from the overall gender distribution of participants. Unsurprisingly, most participants identified as being under personal stress (66.6\%), which was the only significant, positively correlated factor in any of the models ($\beta$ = 1.13, p = 0.036). 

\subsubsection{Career} 

A total of 9 game experiences were reported to lead to career outcomes. While the motivation and game use models revealed no significant predictors (see Figure \ref{career_results}), the most common eudaimonic experience—deep social bonds with other players (66.6\%)—was a significant predictor in the experience model ($\beta$ = 0.095, p = 0.011). However, as the sample size is quite small, more targeted exploration of sociality's influence on career outcomes is necessary. 

\section{Discussion}
% THE TLDR: While the effects of eudaimonic game experiences resemble those described in the literature, our findings suggest that pathways to the formation of each effect differ.
Our thematic analysis (Section \ref{rq1}) yielded five broad themes categorizing effects of eudaimonic gaming experiences. It is notable that these effects align strongly with those previously described in the literature. Additionally, our analysis of the constituent motivations, game usages, and experiences that led to each effect (Section \ref{rq2}) preliminarily suggests that there are different pathways to the formation of each effect. We discuss how current models of eudaimonic gaming experiences can be integrated with existing models of behavioral effects in games, as well as suggestions for turning differences in effect formation into considerations for design and research. 

\subsection{Adding to current understanding of video game play effects}
% SUMMARY
% - Effects the Same -- Existing Models of How media effects people that connect to 
% - Explain how it connects to the models presented
 
Our surfaced effects align strongly with those presented in previous studies \cite{granic2014benefits, quwaider2019impact, bourgonjon2016players}, and to some extent, with \citet{oliver2021model}'s model of inspiring media. In particular, we saw evidence of `mastery experiences' (e.g., self-efficacy, self-confidence, resilience) \cite{rieger2014media}, as well as all four effect types from \citet{granic2014benefits} and their subeffects from \citet{quwaider2019impact}. Section \ref{rq1if} primarily discusses \textit{emotional} benefits, such as `sources of inspiration and connectivity' and `higher self-esteem' \cite{quwaider2019impact}. Section \ref{rq1learning} shows evidence of perceived \textit{cognitive} effects such as `improved problem-solving skills,' `filtering irrelevant information and allocating resources more effectively,' as well as \textit{motivational} effects such as `increasing intelligence and abilities,' `resilience in the face of failure,' and `using failure as motivational tool' \cite{quwaider2019impact}. Section \ref{rq1sc} elaborates on social benefits, such as `rapidly learning social skills and prosocial behaviors' and `increasing group organization and leadership skills' \cite{quwaider2019impact}. Section \ref{rq1health} coveres subthemes of `mood management' and `enhancing positive feelings,' both through physical and mental means \cite{quwaider2019impact}. Finally, Section \ref{rq1career} presents limited evidence for games being `effective for training new behaviors' and `leading to lasting educational success' \cite{quwaider2019impact}. 

This is promising for game researchers utilizing eudaimonic frameworks in their work. It is likely possible to connect and validate pathways from \citet{Possler_2024}'s theorized model of eudaimonic gaming experience formation to models of broader video game effects, such as \citet{granic2014benefits}'s, as well as to models of positive media effects more generally \cite{oliver2021model}. Additionally, drawing from effects elaborated on in the arts may be valuable; for example, \citet{bourgonjon2016players}'s categorization of video game effects, which draws from \citet{belfiore2008social}'s types of cultural rhetoric, also aligns strongly with our findings (e.g., self-development, personal well-being, civilization/worldview). Another promising area of future exploration is connecting eudaimonic gaming experiences with meaningful learning experiences. We saw several instances of meaningful experiences leading to informal learning as described in \citet{iacovides2014gaming}'s model, on a \textit{skill level} (namely, psycho-motor, cognitive, social, numeracy, literacy, and technical skills) as well as a personal level (e.g., cultural and emotional development, career influence).

There remain areas where the existing model of eudaimonic gaming experiences could be expanded upon. The literature has recently expanded upon conceptions of relatedness; for example, the BANGS \cite{Ballou_Denisova_Ryan_Rigby_Deterding_2024}, an update to the PENS \cite{Ryan_Rigby_Przybylski_2006}, encompasses relatedness to characters, as suggested by \citet{Tyack_Wyeth_2017}. However, our findings serve as additional evidence to further incorporate \citet{Tyack_Wyeth_2017}'s suggestions to explore how relatedness to game culture may factor into need satisfaction. For example, several mentions were made to fan communities, which can be powerful venues for informal learning \cite{evans2017more}. Exploring how this "group habitus" \cite{Tyack_Wyeth_2017} may lead to eudaimonic effects, especially after the initial gameplay, could be valuable. There is also an opportunity to further elaborate on the model using OIT's framing for meaningful internalization \cite{pelletier2023organismic}. In particular, OIT may help frame effects such as identity formation and shifts in one's worldview through engagement with parable narratives \cite{Lu_Moller_2024, przybylski2012ideal}, which were predominant in our dataset. Finally, escapism is generally considered a \textit{motivation} \cite{iacovides2019role, holl2023motivation} rather than an \textit{experience} (in the framing of \citet{Possler_2024}'s model). However, for some participants, `healthy escapism' \cite{kosa2020four} appeared to be the meaningful experience, particularly in times of personal stress or crisis. \citet{Cole_Gillies_2022}'s suggestion to combine \citet{stenseng2012activity}'s \textit{self-expansive} escapism with Hartmann's conception of media consumption \cite{hartmann2013media} in the context of eudaimonic gaming experiences may be a valuable extension to \citet{Possler_2024}'s model, as it frames escapism as the motivation with psychological growth as the experience. 

\subsection{Supporting incidental vs. intentional eudaimonia} Overall, our findings in Section \ref{descr_exp} mirrored those in \citet{bowman2024excited}: that ``deeper emotional reactions to video games are not so much anticipated as they are experienced during and after gameplay.'' Additionally, the majority of participants considered themselves to be in normal life circumstances. This suggests that generally, eudaimonic experiences are \textit{incidental}. It is unlikely, for example, that participants who experienced career-related outcomes started playing the game with this in mind, but engaging with other players and game-adjacent online communities helped them develop the requisite skills to effect change. Similarly, participants who reported positive inward-focused effects likely did not expect to self-reflect or identify with characters so significantly. 

However, as discussed in \citet{iacovides2019role} and supported by our results in Section \ref{rq2health}, it is likely that players in stressful circumstances \textit{intentionally} seek meaningful experiences that may result in positive effects for their lives. This is further supported by literature exploring gaming as a coping mechanism \cite{kosa2020four}. We see alignment with \citet{boldi2022commercial}'s findings with regard to players ``appropriating'' games 
in crisis situations; that is, games ``enabl[ed] the ascription of meaning to virtual spaces,'' and players found ways to imbue games with their own meanings to leverage them as support mechanisms. While participants may have initially wished to engage in self-suppressive escapism (associated with avoidant coping) by playing in times of personal stress, having a eudaimonic experience resulted in the overall effect of self-expansive escapism, giving participants \textit{approaches} with which to cope \cite{stenseng2012activity}. It is likely that our findings around how being emotionally moved or challenged and gaining insight into the narrative may spur the inward-focused reflection necessary to progress from self-suppressive to self-expansive escapism. These findings are consistent with narrative theory; although players may not always share the game character's views, transportation into the story leads players to play in accordance with the game character's narrative arc \cite{Happ_Melzer_Steffgen_2013,Yin_Xiao_2022} by means of ``position[ing] themselves within the mental models of the story'' \cite{Busselle_Bilandzic_2008}, which can have the resulting effect of changing players' own mental models \cite{igartua2012changing}. While previous research emphasizes the gains in intrinsic motivation and internalization that result from players being allowed to model their `ideal selves' in games \cite{przybylski2012ideal}, the predominance of parable narratives in our findings would suggest that it may be valuable to struggling players to play as characters growing from their own suboptimal circumstances, challenging players' own views and providing them with approaches to create change in their own lives \cite{Lu_Moller_2024}. Although our findings are preliminary, we believe that this transformation from avoidant to approach-based coping, evocative of `upward spirals' referenced in cognitive-behavioral therapy \cite{fredrickson2002positive}, merits further theorization in both clinical and HCI settings. 

There is also an opportunity to design technical solutions around intentionality; for example, it may be valuable to more directly surface how certain games have affected others meaningfully through game reviews on platforms such as Steam or Reddit. This would also be applicable to self-tracking platforms used for narrative and/or interactive media generally, such as Goodreads, Spotify, or Letterboxd, existing solutions for sharing and reflecting on media. Taking a uses and gratifications perspective, such as \citet{lukoff2018makes}'s investigation of eudaimonic usages of smartphones, may help further elaborate on player intentionality for meaningful experiences. Incidental experiences are more difficult to scaffold around, as they are largely centered around personal interpretations, which is often the design intention of the developers \cite{denisova2021whatever}. Our results largely align with \citet{mekler2018game} in that we observed ``lingering and memorable gaming experiences promot[ing] critical reflection'' \cite{marsh2013lingering, khaled2018questions}, but also that ``these sorts of experiences do not guarantee it,'' as only some participants had higher-level reflective outcomes. Further investigation into the pathway from experience to critical reflection and transformation over the course of several years is likely necessary, and this may provide insight as to whether such reflection can actually be prompted.

\subsection{Pathways to informal learning and career development}
Learning outcomes primarily seemed to center on exploring the content of the game narrative, e.g., mythology in mythology-heavy games, or recreating different aesthetic components of the game, e.g., music and art. We viewed this as an example of informal \cite{iacovides2014gaming} or interest-driven learning \cite{gee2013games} through games, rather than game-based learning, which focuses on ``design[ing] games to maximize learning'' \cite{plass2020handbook}. Hobbies can be productive venues for deep engagement and learning \cite{azevedo2013tailored}, and further exploration into how eudaimonic experiences may prompt instances of learning would be valuable \cite{squire2007open}. Engaging eudaimonic frameworks with existing models of informal learning in games (e.g., \cite{iacovides2014gaming}) may also shed some insight into what forms of personal meaning may mediate different learning outcomes. For example, while we only witnessed a few instances of career change, it is perhaps notable that a large percentage of these outcomes appeared to be social experiences, in contrast to learning outcomes, which were negatively correlated to social bonds. This may suggest that community involvement, as well as skills and encouragement received from engaging in the community (e.g., \cite{lee2022community}), were key to transforming the game experience into something beyond merely educational (also suggested by \citet{iacovides2014gaming}). 

\subsection{Guiding young people from experience to effect} Parental behavior surrounding play affects meaningful internalization \cite{bradt2024does, van2019parents}, which likely affects the formation of eudaimonic effects.
As many of our participants were children or adolescents at the time of play, creating methods for parents and educators to engage in reflective discussions with children around games that are meaningful to them may guide them from experience to effect. Indeed, `critically reading' games, similarly to how students engage with books in English literature classes, has been suggested but underexplored over the last two decades \cite{beavis2005reading, apperley2013model, mctigue2019getting}. Furthermore, it is notable that only one of our participants explicitly mentioned discussing their meaningful experience with their parents. Common Sense Media reported parental co-use of console or tablet games as lower than traditional viewing media such as television and YouTube \cite{parents_2025}. This is an opportunity to promote more active engagement between parents, educators, and children regarding these experiences. For research, this could mean emphasis on the potential for eudaimonic effects in models of parental technology acceptance for video games (e.g., \cite{bassiouni2016video}), or using principles from joint media engagement \cite{ewin2021impact} to design activities around promoting meaningful internalization. For practice, this could mean expanding on potential eudaimonic elements in technology recommendation hubs such as Common Sense Media. Additionally, there may be opportunities to design for mental health practitioners and clinicians looking to engage their adolescent patients in conversations around self-discovery and self-acceptance through their interests in games. 

\section{Limitations}
There are several limitations to this work. Many of the participants were sourced from Prolific, which limits geographic and demographic diversity and therefore makes it difficult to understand the possible scope of meaningful experiences. Although the aesthetic component of `motivation' for eudaimonic gaming experiences also involves player interaction with the soundtrack and controller haptics, we could not find an adequate framework with which to classify these elements. While we did not fit our model to account for interaction effects, future research could explore moderated relationships between specific components of \citet{Possler_2024}'s process model using larger quantitative samples or alternative modeling techniques. Finally, participants primarily discussed positive effects from their gaming experiences; however, it is possible that negative effects can result from these experiences as well. 

\section{Conclusion and Future Work}
\begin{quote}
    ``\textit{I honestly think [that] I would not be this version of myself today without these experiences.}'' - P173
\end{quote}

In this work, we investigated the effects of eudaimonic gaming experiences and how the factors affecting their formation may lead to specific outcomes. We disseminated a survey to gather accounts of these experiences and used a mixed-methods approach heavily grounded in existing theoretical models \cite{Possler_2024} to analyze them. We presented five primary classes of effects and used binomial logistic regression models on our dataset to surface factors that were significant to their formation. We found potential connections between narrative identification and inward-focused effects, stress as a motivator for positive health outcomes, and differences in the pathways leading to interest-driven learning versus career effects. We discuss implications for connecting the study of eudaimonic gaming experiences to existing models of positive video game effects and provide several avenues for potential future exploration. These include investigating how users seek out meaningful gaming experiences for mental health support through platforms such as Reddit, further investigation into how aesthetic elements influence eudaimonia in games, and scale development incorporating all the constituent constructs of the eudaimonic gaming experience. We contribute an empirical understanding of the effects of eudaimonic gaming experiences and provide design considerations for research, parents, and educators.

\begin{acks}
    This research was supported by a doctoral research grant from the Ramey Research Fund and the University of Washington’s Department of Human Centered Design \& Engineering.
\end{acks}
\bibliographystyle{ACM-Reference-Format}
\bibliography{sample-base}

% set formatting for ludography

\bibliographystylegame{ACM-Reference-Format}
\bibliographygame{games}
\end{document}